\documentclass{aa}  

\usepackage{graphicx}
\usepackage{txfonts}
\usepackage{hyperref}
\usepackage{comment}

\usepackage{amsmath}	
\usepackage{amssymb}	
\usepackage{amsbsy}	    
\usepackage{subfig}
\usepackage{caption}
\usepackage{ae,aecompl}
\usepackage[T1]{fontenc}
\usepackage{color}

\newcommand{\msol}{\mathcal{M}_\odot}
\newcommand{\pc}{\mathrm{\ pc}}
\newcommand{\sfe}{\mathrm{SFE_{J}}}
\newcommand{\sfeP}{\mathrm{SFE}_{10}}
\newcommand{\sfeD}{\mathrm{SFE_{10}}}

\newcommand{\eff}{\epsilon_\mathrm{ff}}
\newcommand{\tvr}{t_\mathrm{VR}}
\newcommand{\Myr}{\mathrm{\ Myr}}

\begin{document}

   \title{The bound mass of Dehnen models with centrally
peaked star formation efficiency}

   \titlerunning{The bound mass of Dehnen models with centrally peaked SFE}

   \author{B.~Shukirgaliyev
          \inst{1,2,3}\fnmsep\thanks{Corresponding author}
          \and
          A.~Otebay \inst{3,2,1}
          \and
          M.~Sobolenko \inst{4}
          \and
          M.~Ishchenko \inst{4}
          \and
          O.~Borodina \inst{5}
          \and
          T.~Panamarev \inst{6,2,1}
          \and 
          S.~Myrzakul \inst{7,8,1}
          \and
          M.~Kalambay \inst{3,2,1}
          \and
          A.~Naurzbayeva \inst{3,1,2}
          \and
          E.~Abdikamalov \inst{9,1}
          \and
          E.~Polyachenko\inst{5}
          \and
          S.~Banerjee\inst{10,11}
          \and
          P.~Berczik \inst{12,13,4}
          \and
          R.~Spurzem \inst{12,13,14}
          \and
          A.~Just \inst{13}
          }

   \institute{Energetic Cosmos Laboratory, Nazarbayev University,
              53 Kabanbay Batyr ave., 010000 Nur-sultan, Kazakhstan\\
              \email{bekdaulet.shukirgaliyev@nu.edu.kz}
         \and
             Fesenkov Astrophysical Institute,
             23 Observatory str., 050020 Almaty, Kazakhstan\\
             \email{otebay@aphi.kz}
         \and
             Al-Farabi Kazakh National University,
             71 Al-Farabi ave., 050040 Almaty
         \and
             Main Astronomical Observatory, National Academy of Sciences of Ukraine, 27 Akademika Zabolotnoho St., 03143 Kyiv, Ukraine
        \and
             Institute of Astronomy, Russian Academy of Sciences (INASAN), 48 Pyatnitskaya str., 119017 Moscow, Russian Federation
        \and
            Rudolf Peierls Center for Theoretical Physics, University of Oxford, Parks Road, Oxford, OX1 3PU, United Kingdom
        \and
             Ratbay Myrzakulov Eurasian International Centre for Theoretical Physics, 010009 Nur-Sultan, Kazakhstan
        \and
             Center for Theoretical Physics, Eurasian National University, 010008 Nur-Sultan, Kazakhstan
        \and
            Physics Department, Nazarbayev University, 53 Kabanbay Batyr ave., 010000 Nur-sultan, Kazakhstan
        \and
             Helmholtz-Institut f\"ur Strahlen- und Kernphysik, Nussallee 14-16, D-53115 Bonn, Germany
        \and
              Argelander-Institut f\"ur Astronomie, Auf dem H\"ugel 71, D-53121 Bonn, Germany
        \and
             National Astronomical Observatories and Key Laboratory of Computational Astrophysics, Chinese Academy of Sciences, 20A Datun Rd., Chaoyang District, Beijing 100101, China
        \and
             Astronomisches Rechen-Institut am Zentrum f\"ur Astronomie der Universit\"at Heidelberg, M\"onchhofstrasse 12-14, 69120 Heidelberg, Germany
        \and
             Kavli Institute for Astronomy and Astrophysics at Peking University, 5 Yiheyuan Rd., Haidian District, 100871, Beijing, China
        }

   \date{Received ; accepted }

 
  \abstract
   {Understanding the formation of bound star clusters with low star-formation efficiency (SFE) is very important to know about the star-formation history of galaxies. In N-body models of star cluster evolution after gas expulsion, the Plummer model with outer power law density profile has been used massively.  
   } 
  {
   We study the impact of the density profile slopes on the survivability of the low-SFE star clusters after instantaneous gas expulsion. We compare cases when stellar cluster has Plummer profile and Dehnen profiles with cusp of different slopes at the time of formation.
   }
   { We { determine } the corresponding density profile of the residual gas for a given global SFE, assuming that our model clusters formed with a constant efficiency per free-fall time and hence have shallower density profile of gas than that of stars. We perform direct $N$-body simulations of evolution of clusters initially in virial equilibrium within gas potential after gas removal.}
   {
   We find that the violent relaxation lasts no longer than 20~Myr independently of the density profile power law slopes. Dehnen model clusters survive after violent relaxation with significantly lower SFEs when the global SFE measured within the Jacobi radius or within a half-mass radius. Dehnen $\gamma=0$ model clusters show similar final bound fraction with the Plummer model clusters if global SFE is measured within 10 scale radii. The final bound fraction increases with $\gamma$ values for a given global SFE.
   }
   {
   We conclude that Dehnen clusters better resist the consequences of the violent relaxation followed the instantaneous gas expulsion than the Plummer clusters. Thus the shallower the outer density slope of the low-SFE clusters, the better for their survivability after gas expulsion. Among Dehnen clusters we find that the steeper the inner slope (cusp) the higher the bound mass fraction is retained after violent relaxation for a given global SFE.
   }


   \maketitle
%

\section{Introduction} \label{sec:intro}

Star clusters form in dense gas clumps within molecular clouds \citep{Krumholz+19,Krause+20}. The stellar feedback from the newly born massive stars cleans up the star-formation region from the residual gas within a short timescale before the first supernova explosion (SNe) \citep{Kruijssen+19}. The velocity of the stellar feedback has been estimated to be about $10 \, \mathrm{km\ s^{-1}}$, both from theory and observations \citep{Rahner+19,Grasha+19}. All clusters older than 10 Myr are observed to be gas-free \citep{LL03,Leisawitz1989}. \citet{LL03} concluded that only about 10 percent of the newly formed clusters survive the gas expulsion. {}{ They proposed that most of clusters dissolve early as a consequence of gas expulsion because of their low star-formation efficiency. } \citet{Krumholz+19} refers to the 90 percent weight-loss by star clusters, rather than to the early dissolution of most clusters.

\begin{figure}
    \centering
    \includegraphics[width=\columnwidth]{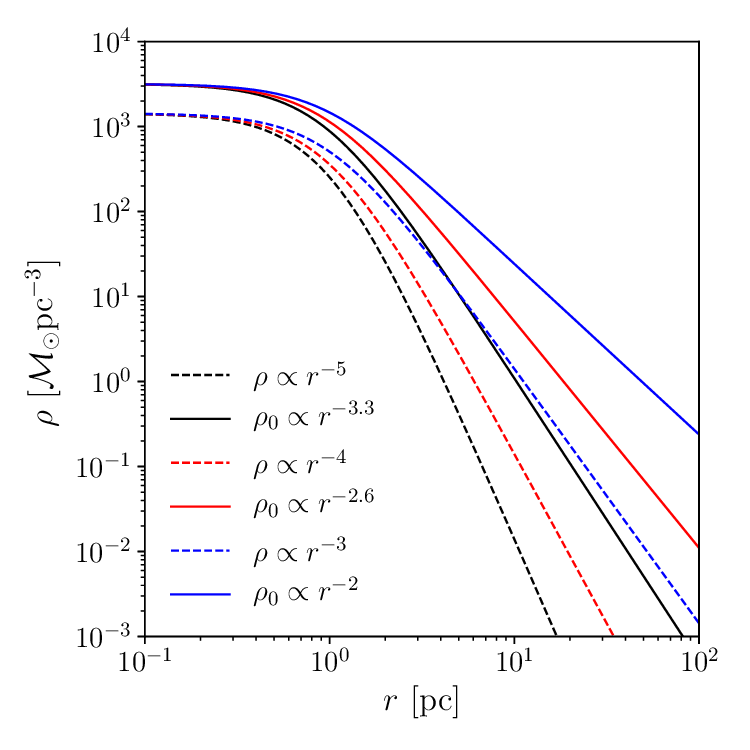}
    \caption{The Plummer-like power law density profiles of stellar clusters for $q=3,4,5$ and of the corresponding initial gas with $p=2,2.6,3.3$. The density profiles of the initial gas were recovered according to the star cluster formation model of \citet{PP13}, alike in \citet{Bek+17}, for $\epsilon_\mathrm{ff}=0.05$, and $t_\mathrm{SF}=3$~Myr. }
    \label{fig:dps}
\end{figure}

{}{ The star-formation efficiency (SFE) -- the fraction of star-forming gas mass converted into stars --  
measured in observed star-forming regions barely reaches 30 percent, being mainly below 20 percent \citep{Higuchi2009,Murray2011,Kainulainen+14}. On the Galactic scale SFE integrated throughout several star-forming regions remain about few percent \citep{Kruijssen+19}.
A large number of works were dedicated to study how star clusters can survive after gas expulsion and subsequent violent relaxation  \citep[][and many others]{Tutukov1978,Hills1980,Lada+1984,Verschueren+1989,Adams2000,GoodwinBastian2006,BK07,Smith+11,LeeGoodwin2016,Bek+17,Brinkmann+17,Farias+18}. 
\citet{BK07} showed that clusters can survive with small SFEs about 10-15 percent, if the residual gas will be expelled gradually within several crossing times. Summarizing the preceded $N$-body simulations they reported that the minimum of 30 percent of star-forming gas should be converted into stars to survive the instantaneous gas expulsion as a bound cluster.
Under SFE we understand here the {\it total} SFE
\begin{equation}\label{eq:SFE_tot}
    \mathrm{SFE_{tot}}=\frac{M_\star}{M_0}=\frac{M_\star}{M_\star+M_\mathrm{gas}},
\end{equation}
where $M_\star$ is the total mass of stars formed in the clump before gas expulsion, $M_\mathrm{gas}$ is the residual gas mass at the time of gas expulsion, and the total initial mass of the star-forming clump is $M_0=M_\star+M_\mathrm{gas}$.

\citet{BK07} considered the Plummer model \citep{Plummer_1911} star clusters in virial equilibrium within total gravitational potential of stars and residual gas immediately before gas expulsion. The density profiles of stars and gas have the same shapes -- { i.e. Plummer profiles with the same scale radius, $a_\mathrm{P}$, but different masses -- in their study. Therefore the SFE is radially constant within their model embedded clusters. Here we bring the form of the Plummer profile for the sake of clarity
\begin{equation}\label{eq:rho_P}
    \rho_\mathrm{P}(r) = \frac{3M_\star}{4\pi a_\mathrm{P}^3} \left(1+\frac{r^2}{a_\mathrm{P}^2}\right)^{-5/2}.
\end{equation}
}

\citet{GoodwinBastian2006} \citep[see also][]{Goodwin2009} introduced the {\it effective} SFE (eSFE) defined based on the dynamical state of the cluster immediately after gas expulsion
\begin{equation}\label{eq:eSFE}
    \mathrm{eSFE} = \frac{1}{2Q}. 
\end{equation}
{ Here the virial ratio, $Q$, is the ratio of the total stellar kinetic energy, $K$, to absolute value of the total stellar potential energy, $W$,
\begin{equation}
    Q=\frac{K}{|W|},
\end{equation}
and the virial equilibrium is defined at $Q=1/2$.}
They concluded that star clusters can survive instantaneous gas expulsion if $\mathrm{eSFE}>0.30$. For model clusters of \citet{BK07} the eSFE is equivalent to the total SFE{, because gas and stars follow the same density profile and stars are in virial equilibrium in the total gravitational potential. } However, the eSFE can be different from total SFE, depending on the cluster virial state { before gas expulsion} \citep{Goodwin2009,LeeGoodwin2016}. Thus \citet{Goodwin2009} discussed the survivability of star clusters which are not in virial equilibrium within total gravitational potential of stars and gas before the gas expulsion. If the gas-embedded clusters are { sub-virial (i.e. $2K<|W|$)} before gas expulsion, then their eSFE can be larger than their total SFE, and thus survive the instantaneous gas expulsion with a low total SFE \citep[][]{Verschueren+1989,Verschueren1990,LeeGoodwin2016,Li+2019}.

\citet{Adams2000} showed semi-analytically that low-SFE clusters can survive instantaneous gas expulsion, if the density profile of stars has a steeper outer slope than that of the residual gas. In this case the local SFE $(\mathrm{SFE_{loc}}(r))$ --  SFE measured locally at arbitrary location within the cluster is not radially constant. We can define the $\mathrm{SFE_{loc}}(r)$ as the ratio of the stellar density to the total density within a given region
\begin{equation}\label{eq:SFE_loc}
    \mathrm{SFE_{loc}}(r) = \frac{\rho_\star(r)}{\rho_\star(r)+\rho_\mathrm{gas}(r)}= \frac{\rho_\star(r)}{\rho_0(r)},
\end{equation} 
where $\rho_\star(r)$, $\rho_\mathrm{gas}(r)$ and $\rho_0(r)$ are local density of stars, unprocessed gas and the total initial starless gas, respectively. The cumulative SFE   
\begin{equation}\label{eq:SFE_r}
    \mathrm{SFE_r}(r) = \frac{M_\star(<r)}{M_\star(<r)+M_\mathrm{gas}(<r)}, 
\end{equation}
also varies radially, decreasing with radius, since the cumulative mass of the residual gas grows faster with radius than that of stars. 
Especially, in the case of \citet{Adams2000}, when the residual gas density is $\rho_\mathrm{gas}\propto r^{-2}$, $\mathrm{SFE_r}(r)$ continuously decreases with radius, because the residual gas mass diverges. 
Therefore \citet{Adams2000} introduced the outer truncation radius, where stellar density $\rho_\star$ becomes zero in his models, to measure SFE. However, depending on the density profile applied for stars, this kind of truncation radius can be as large as $+\infty$ (e.g. for the Plummer model).

\citet{Smith+11} proposed the formation of clusters through hierarchical merger of sub-structured clusters within different residual gas backgrounds. They considered cluster models with $\mathrm{SFE_{tot}}=0.20$ where stars are distributed within fractal sub-clusters \citep{GoodwinWhitworth2004} or clumpy Plummer spheres, and arbitrarily chosen different gaseous backgrounds (from the Plummer to homogeneous spheres).

\citet{Smith+11} concluded that the key parameter showing whether cluster survives the instantaneous gas expulsion is not the total SFE, but the {\it local stellar fraction} (LSF), that is cumulative SFE measured within the half-mass radius of the stellar component of the embedded cluster, $r_\mathrm{h}$ at the onset of gas expulsion
\begin{equation}\label{eq:LSF}
    \mathrm{LSF} = \frac{M_\star(<r_\mathrm{h})}{M_\star(<r_\mathrm{h})+M_\mathrm{gas}(<r_\mathrm{h})}\equiv \mathrm{SFE_r}(r_\mathrm{h}).
\end{equation}
Cluster formation through hierarchical merger of sub-clusters also appears in hydro-dynamical simulations starting from initially homogeneous sphere of molecular gas \citep[e.g.][]{Wall+2019,Li+2019,Wall+2020,starforge,Fukushima+Yajima2021}. However, \citet{Chen+2021} have showed recently that initially centrally-concentrated gas clouds (i.e. with power density profile) tend to form massive central cluster which grows in mass through the accretion of gas around. Additionally, they noticed that the steeper the power law density profile the more massive is the central cluster.  

\citet{Bek+17} performed series of $N$-body simulations of bound cluster formation after instantaneous gas expulsion with physically motivated star-formation conditions of \citet{PP13}. The local-density driven clustered star formation model, proposed by \citet{PP13}, assumes that stars form in centrally-concentrated spherically-symmetric dense gas clump  with a constant SFE per free-fall time $(\epsilon_\mathrm{ff}=\mathrm{const})$. { Here the free-fall time is defined as
\begin{equation}
    \tau_\mathrm{ff}(r) = \sqrt{\frac{3\pi}{32G\rho_\mathrm{gas}(r)}},
\end{equation} 
where $G$ is the gravitational constant.}
As a consequence, the inner dense gas produces more stars than the outer diffuse gas within a given physical time of star-formation (SF) duration, $t_\mathrm{SF}$. Thus stars have steeper { (power law)
} density profile than the initial starless gas and the residual gas before gas expulsion \citep{PP13,Bek+17}. 

Assuming that the \citeauthor{Plummer_1911} star clusters formed with $\epsilon_\mathrm{ff}=\mathrm{const}$ during $t_\mathrm{SF}$,
\citet{Bek+17} reconstructed the corresponding density profiles of the residual gas before gas expulsion. They show that the product of $\epsilon_\mathrm{ff} \times t_\mathrm{SF}$ determines the cluster's global SFE.  \citet{Bek+17} defined the global SFE in their models as the cumulative SFE measured within 10 Plummer scale radii ($10 a_\mathrm{P}$), where over 98 percent of the stellar mass resides at the time of instantaneous gas expulsion
\begin{equation}\label{eq:SFE_gl}
    \mathrm{SFE_{gl}}=\mathrm{SFE_r}(10a_\mathrm{P}).
\end{equation}
They showed that their model clusters have a centrally peaked SFE-profile \citep[see e.g. Fig. 2 of ][]{BekPhDT18}. From the results of $N$-body simulations, they concluded that the minimum global SFE needed to survive as a bound cluster is $\mathrm{SFE_{gl}}=0.15$ for their model clusters, independent of the cluster stellar mass \citep{Bek+18} and of the impact of the Galactic tidal field \citep{Bek+19,Bek+20}. 

In the observed nearby star-forming regions, gas and stars do not follow the same density profiles. \citet{Gutermuth+2011} reported about the power law correlation {
\begin{equation}
    \Sigma_\star \propto \Sigma_\mathrm{gas}^2
\end{equation}
between the local surface densities of young stellar objects, $\Sigma_\star$, and the column density of gas, $\Sigma_\mathrm{gas}$, in 8 nearby star-forming regions \citep[see also the recent results by ][]{Pokhrel+20}. This corresponds to an increase in SFE with increasing gas density meaning that stellar density profile has a steeper power law slope than that of gas. \citet{PP13} explained that such a correlation is the consequence of the star formation taking place with a constant efficiency per free-fall time with their local-density driven clustered star formation model. 
In the state-of-the-art hybrid hydro-dynamical/$N$-body simulations of clustered star formation the residual gas and newly formed stars follow power law density profiles with indexes of about 2 and 3, respectively \citep{Li+2019,Sirius2}.
} 
In fact, \citet{PP13} noticed that the outer slope of the power law density profile of the newly formed stellar cluster will be steeper by a factor of 1.5 at most than that of the initial starless gas according to their model. } That is, if the initial starless gas has a power law profile $\rho_0 \propto r^{-p}$ with an index of $p=2$, then stars would follow the power law profile $\rho_\star\propto r^{-q}$ with index of $q\leq 3p/2 = 3$ \citep{PP13}. { In case of \citet{Bek+17} the formed cluster follows the Plummer density profile, i.e. $q=5$ at the outer part, and the density profile of the initial star-forming gas at the outer part follows the power-law with ${p\approx 3.3}$.} { In Fig.~\ref{fig:dps} we demonstrate this relation between the slopes of power law densities for a few fiducial cases of the Plummer-like stellar density profiles $\rho_\star\propto(1+r^2/a_\star^2)^{-q/2}$ (see Eq.~\ref{eq:rho_P}) with $q=3,4,5$ (for $a_\star=1\ \mathrm{pc}$ and $M_\star=6000\msol$), and the initial gas power law density profiles $\rho_0\propto(1+r^2/a^2)^{-p/2}$ with $p=2,2.6,3.3$, respectively. The initial gas profile is $\rho_0=\rho_\star+\rho_\mathrm{gas}(\epsilon_\mathrm{ff}t_\mathrm{SF})$, where $\rho_\mathrm{gas}$ is the recovered density profile of the residual gas after time $t_\mathrm{SF}=3$~Myr since the onset of the star formation with $\eff=0.05$.}
{ 
The power law density slope of the observed gas clumps commonly varies in the range of $1\leq p\leq 2$ \citep{Lada+2010, Kainulainen+14, PP2020}. However, \citet{Schneider+15} reported about molecular gas clumps having a steep power law density slope up to $p<4$. That is any of the three fiducial models we show in Fig.~\ref{fig:dps} may exist according to observations.

Almost all studies investigating the evolution of star clusters using $N$-body simulations consider the Plummer profile to represent the stellar density distribution before gas expulsion \citep[][]{BK07, LeeGoodwin2016,Bek+17}. 
However, the commonly observed }gaseous clumps have rather shallow density profiles than that of the residual gas in case of the Plummer profile ($<p=3.3$). This arises the following questions: Is it possible that stellar clusters have shallower slope of density profile than that of the Plummer profile at the time of formation before gas expulsion? And if so, does it help such clusters to survive the instantaneous gas expulsion better than in case of previously considered Plummer model? 

{ \citet{Li+2019} hydro-dynamical followed by $N$-body simulations show that star clusters with a shallow ($2<q<3.5$) power law density profile can survive the stellar feedback driven gas expulsion with total SFEs below 10 percent. \citet{sirius3} summarizing the results from their hybrid hydro-dynamical/$N$-body simulations of cluster formation stated that considering the simplified models of star formation and instantaneous gas expulsion is sufficient to study cluster mass function and dynamical evolution of star clusters after gas expulsion. Because, the hybrid hydro-dynamical/$N$-body simulations are expensive computationally, they can be limited to a few models or to the low-mass or low-resolution cases \citep{Li+2019}.
 So } we decide to continue to study the star cluster dynamical evolution after gas expulsion with our methods developed in \citet{Bek+17}. 
{ Since the Plummer model has too steep power law density to represent the stellar component of the commonly observed embedded clusters, we propose a new set of numerical experiments using the family of the \citet{dehnen_family_1993} model density profiles (with an outer slope of $q=4$). The Dehnen model clusters have shallower power density outer slope than that of the Plummer, but have finite mass in contrast to the case of power density of $q=3$. Also the power-law index of density profile of the star-forming gas expected to be $p \geq 8/3 \approx 2.6$, which is still in the range of the observed values for molecular clumps \citep{Schneider+15}.}


In this work we study the impact of the outer and inner density slopes on the survivability of star clusters after instantaneous gas expulsion. { We consider the instantaneous gas expulsion since it is the worst scenario for the cluster survival after gas expulsion, thus represents the lower-limit of the cluster survivability. The gradual gas expulsion allows clusters to keep more stars bound after violent relaxation than the instantaneous one \citep{GeyerBurkert2001,BK07,Brinkmann+17}. It is still unclear how does the gas expulsion happens in real systems \citep{Krumholz+Matzner2009,sirius3}. 
Nevertheless, the gas expulsion phase, which is mostly driven by the most massive O-B stars, does not last much longer than the free-fall time of the star-forming clump \citep{Li+2019}, which results not too different from the instantaneous gas expulsion \citep{BK07}. Also, since only the massive stars are responsible for gas expulsion, it is enough if a few most massive stars switch on their stellar wind within short time interval to start drive gas out of the cluster. 
}


{\it}{
In section \ref{sec:models} we describe our Dehnen model clusters in comparison with the Plummer model case. Also we discuss about measuring SFE with different methods and how they are done for our models. Then we present the main results in section \ref{sec:res}, i.e. the bound mass evolution and the survivability of model clusters. In section \ref{sec:conc} we provide the discussions and summarize our new results.
}

\section{Methods and models}\label{sec:models}

\subsection{Models of stellar cluster}

{ We build star cluster models with different formation conditions (i.e. different SFE), where the stellar component of the embedded cluster has identical properties (e.g. mass, density profile, size). That is the properties (e.g. density profile, mass) of the initial/residual gas are reconstructed assuming that our model clusters were formed according to the local-density-driven cluster formation model of \citet{PP13} for a given global SFE.
This ``upside-down'' approach compared to \citet{PP13} is needed for us to be able to compare star cluster models with different SFE among each other. That is, if we would start from the molecular clump with a given mass and size, the newly formed star clusters with different SFEs would differ also on sizes, masses and density profiles. That will make them incomparable to each other. In the other way around, this method will allow us to explore the long-term evolution of star clusters until their dissolution in the tidal field of the host galaxy taking into account the formation conditions in a wide variety of the initial conditions \citep[see also ][]{BekPhDT18}.
}

\begin{figure}[ht!]
    \includegraphics[width=\columnwidth]{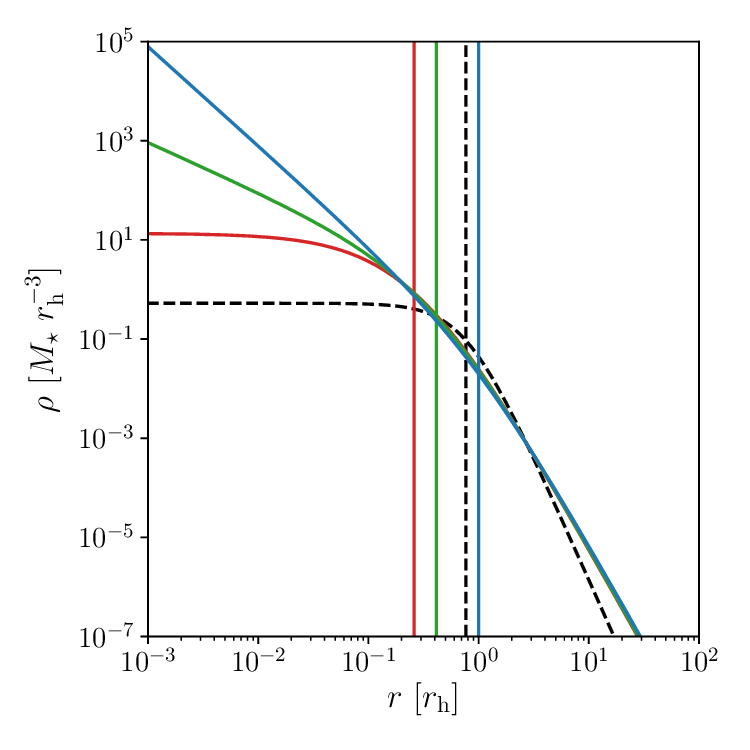}
    \caption{Volume density profiles of star clusters with equal masses and half-mass radii corresponding to the Dehnen models ($\gamma=0,1,2$ in red, green and blue solid lines, respectively), and the Plummer model (black dashed line). The vertical lines show the corresponding scale radii, $a_\mathrm{D}$ and $a_\mathrm{P}$ with respective colors. \label{fig:den-prof}}
\end{figure}

\begin{figure}[ht!]
    \includegraphics[width=\columnwidth]{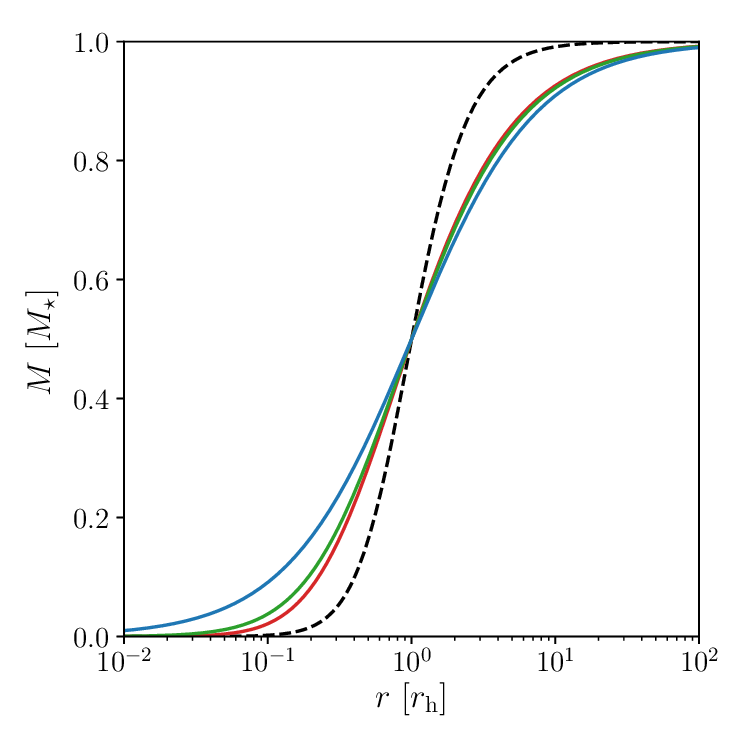}
    \caption{Cumulative mass profiles of star clusters with equal masses and half-mass radii corresponding to the Dehnen models with($\gamma=0,1,2$ in red, green and blue solid lines, respectively), and the Plummer model (black dashed line).
    \label{fig:mass-prof}}
\end{figure}

\citet{dehnen_family_1993} introduced the family of two-power-density spherical models. This models originally intended to describe the spatial distributions of stars in galaxies, and never has been used to describe star clusters due to its large outer radius. {}{ We bring the density profile expression of the Dehnen model here, for the sake of clarity
\begin{equation}\label{eq:rho_D}
    \rho_D(r) = \frac{(3-\gamma)M_\star}{4\pi} \frac{a_\mathrm{D}}{r^\gamma (r+a_\mathrm{D})^{4-\gamma} },
\end{equation}
where $M_\star$ is the total stellar mass, $a_\mathrm{D}$ is the Dehnen scaling radius and $\gamma$ describes the inner power-law profile of the family of Dehnen models ($0\leq \gamma<3$). 
} 
Figure \ref{fig:den-prof} 
shows the comparison of these density profiles for $\gamma=0,1,2$ (red, green and blue solid lines) and the Plummer model (black dashed line) clusters, while Fig.~\ref{fig:mass-prof} 
demonstrates their cumulative mass distributions. We assume that clusters have identical stellar mass, $M_\star$, and half-mass radius, $r_\mathrm{h}$. Since we equate masses and half-mass radii of the our model clusters, their scale radii become different from each other (see vertical lines in Fig.~\ref{fig:den-prof}). 
The relations between half-mass and scale radii for the Dehnen and the Plummer models are given below:
\begin{align}
    r_\mathrm{h} &= a_\mathrm{D}\left(2^{1/(3-\gamma)}-1\right)^{-1}, \label{eq:rh_D} \\
    r_\mathrm{h} &\approx 3.84a_\mathrm{D}  &(\mathrm{ for }\ \gamma=0),\nonumber \\
    r_\mathrm{h} &\approx 2.41a_\mathrm{D}  &(\mathrm{ for }\ \gamma=1),\nonumber \\
    r_\mathrm{h} &= a_\mathrm{D} &(\mathrm{ for }\ \gamma=2),\nonumber 
\end{align}
and 
\begin{equation}\label{eq:rh_P}
    r_\mathrm{h} = a_\mathrm{P}\left(2^{2/3}-1\right)^{-1/2}\approx 1.3a_\mathrm{P}.
\end{equation}
The units of distance and mass in Fig.~\ref{fig:den-prof} are normalized to the cluster half-mass radius and stellar mass, thus densities are presented in the corresponding units of $\left[M_\star r_\mathrm{h}^{-3}\right]$.  Note, that for equal scale radii, the central densities of the Dehnen $\gamma=0$ and the Plummer models are equal, but then their half-mass radii will be different. The half-mass radius is used to describe the cluster size in the majority of star cluster studies. Thus to be consistent with them we also choose the half-mass radius of star clusters to represent their sizes.
In Dehnen models, the transition from the inner power-law to the outer power-law is smoother compared to that of the Plummer model. 
From Fig.~\ref{fig:mass-prof} we can see that the Dehnen model clusters contain about 90 percent of their mass within the sphere of radius larger than $10r_\mathrm{h}$. The Plummer model is quite compact in this sense and reaches up to 90 percent of its mass already within $3r_\mathrm{h}$. Although, the Dehnen clusters have more compact and denser core than the Plummer cluster.

{
\citet{Sirius2} showed \citep[also see][]{Li+2019} that star clusters do not have a core during the gas embedded phase. That motivates us to consider the Dehnen models with $\gamma>0$ (i.e. with a cusp) in our study too. }

\subsection{Recovering the residual gas}

As mentioned before, we assume that in our model clusters star-formation happens with a constant SFE per free-fall time ($\epsilon_\mathrm{ff}=\mathrm{const}$) according to \citet{PP13}. Thus following \citet{Bek+17} we can recover the density profile of the residual gas clump before gas expulsion, for a given stellar density profile, $\rho_\star(r)$, SFE per free-fall time, $\epsilon_\mathrm{ff}$, and star-formation duration, $t_\mathrm{SF}$. For the sake of clarity we bring these expressions \citep[Eqs. A1-A7]{Bek+17} here
\begin{equation}
\label{eq:rhogas1}
    {\rho_\mathrm{gas}(r)}=\frac{1}{k^2}-\frac{\rho_\star(r)}{2} - \frac{1}{2} \sqrt{K_2 + \frac{8}{k^6 K_1}} + K_1,
\end{equation}
\begin{equation}
k = \sqrt{ \frac{8G}{3\pi} } \epsilon_\mathrm{ff} t_\mathrm{SF}.
\end{equation}
\begin{equation}
\alpha = k^{4}\rho_\star^2,
\end{equation}
\begin{equation}
K_0=\sqrt[3]{\alpha^3+36 \alpha^2+216 \alpha +24 \alpha \sqrt{3 \left(\alpha+27\right)}},
\end{equation}
\begin{equation}
K_1=\sqrt{\frac{\alpha^2+\alpha(K_0+24)+K_0 (K_0+12)}{12 k^4 K_0}},
\end{equation}
\begin{equation}
\label{eq:rhogaslast}
K_2=\frac{\left(\alpha-K_0+24\right) \left(K_0-\alpha\right)}{3 k^4 K_0}.
\end{equation}
{}{ The solution of \citet{Bek+17} allows for the stellar density profile to be any centrally-concentrated function. Since gas density depends on the stellar density locally, the solution is also valid for clumpy structure of stellar cluster.}

To account for the residual gas potential in the case of Dehnen model clusters, we have developed a new acceleration { (read as external potential\footnote{ In the \texttt{NEMO/falcON} package term acceleration plug-in is used for a plug-in accounting for additional acceleration from the background potential when generating the initial conditions of $N$-body system in virial equilibrium within some external gravitational potential. })} plug-in \texttt{GPDehnen}\footnote{Publicly available in the github portal:\\ \href{https://github.com/BS-astronomer/GasPotential}{https://github.com/BS-astronomer/GasPotential}} to the \textsc{mhkhalo} program \citep{McmillanDehnen07} from the \texttt{NEMO/falcON} package \citep{Teuben1995,Dehnen2000,Dehnen2002}, based on previous \texttt{GasPotential} acceleration plug-in \citep{Bek+17,BekPhDT18}. Together with \texttt{GPDehnen} acceleration plug-in, \textsc{mkhalo} program generates a single-mass $N$-body system, distributed in position-velocity space by the Dehnen model in virial equilibrium within the total gravitational potential of stars and gas. \texttt{GPDehnen} acceleration plug-in requires 4 parameters: $\epsilon_\mathrm{ff}$, $t_\mathrm{SF}$, $\gamma$ and $a_\mathrm{D}$ to reproduce the gravitational potential of the residual gas. The stellar mass is set to the unity in $N$-body units. {}{ We also have developed another method to generate our special initial conditions with \textsc{Agama} code \citep{agama}.  Both methods work very well and the generated $N$-body systems are in virial equilibrium within the external potential of the residual gas recovered by Eq.~\ref{eq:rhogas1}. } { One can re-assign the masses of the individual particles according to the chosen initial mass function (IMF) keeping the total mass constant. That will introduce some perturbations locally, but won't change the total virial ratio of the generated $N$-body system. Therefore it stays in virial equilibrium with the external potential after even after introducing the IMF. }

{}{ In this study, similar to previous studies, we adopt SFE per free-fall time $\epsilon_\mathrm{ff}=0.05$. However, as it is clear from Eqs (\ref{eq:rhogas1}-\ref{eq:rhogaslast}), the density profile of the residual gas depends on the product of $\epsilon_\mathrm{ff}$ and $t_\mathrm{SF}$ rather than their individual values. Therefore any values of $\epsilon_\mathrm{ff}$ and $t_\mathrm{SF}$ can be assumed without changing the global SFE while their product is preserved. From observations of star-forming regions, SFE per free-fall time has been estimated to be about $\epsilon_\mathrm{ff}\approx 0.01$. The values of $\log \epsilon_\mathrm{ff}$ varies from $-2.5$ to $-1.3$ depending on the method of measurements \citep[see Fig. 10 in][]{Krumholz+19}. 
If we want SFE per free-fall time to be $\epsilon_\mathrm{ff}=0.01$, the only changes to be applied to the adopted values of $t_\mathrm{SF},$ which becomes 5 times longer than in case of $\epsilon_\mathrm{ff}=0.05$. It will not change anything else for the subsequent $N$-body simulations going after instantaneous gas expulsion.}

\subsection{Measuring SFE}

Our model clusters have centrally peaked SFE profiles, due to the fact that stars have steeper density profile than gas, as a consequence of star-formation happening with a constant efficiency per free-fall time. That means, as we noticed before,  SFE decreases with increasing the radius of measurement.
In case of the Plummer model, the cumulative SFE converges to the total SFE at the infinity, because the residual gas mass converges too ($p\approx 3.3$). In case of Dehnen models, the global SFE vanishes at infinite radius. This is caused by diverging mass of the residual gas, which has a power-law density profile with index $(p\approx 2.6 < 3)$. Therefore we need to define robustly the cluster outer radius to measure the value of SFE representing the global SFE of model clusters. { Even in the case of the Plummer model we had to determine the global SFE within some outer radius to be consistent with other studies. } In case of real star-forming regions, one cannot increase infinitely the outer radius, which then starts to account for the neighboring star-formation regions, since molecular clouds usually contain few or several star-forming regions next to each other. There is no universal definition of the cluster outer radius to measure the global SFE can be found neither from theory nor from observations. This also puts some uncertainties to comparison of models and observations. 

\begin{figure}
    \centering
    \includegraphics[width=\hsize]{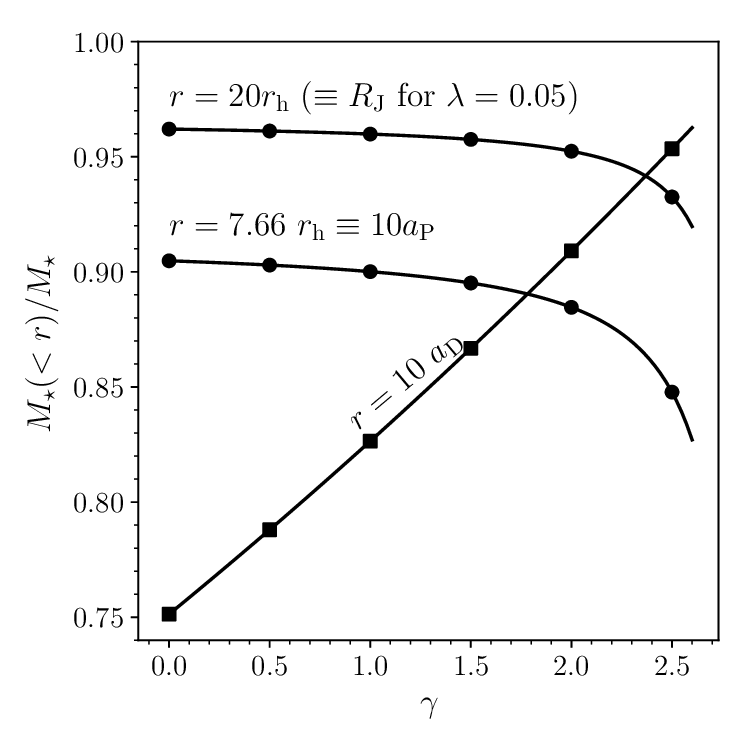}
    \caption{Enclosed mass fraction of the Dehnen model clusters within different radii ($R_\mathrm{J}\equiv20r_\mathrm{h},10a_\mathrm{P}\equiv 7.66r_\mathrm{h},\mathrm{\ and\ }10a_\mathrm{D}$) as a function of $\gamma$.}
    \label{fig:massfrac}
\end{figure}

In previous studies \citep{Bek+17}, for the Plummer model clusters the outer radius $R_\mathrm{out}=10a_\mathrm{P}=7.66r_\mathrm{h}$ has been used to measure the global SFE, which then used to parameterize the set of models. However, neither $R_\mathrm{out}=10a_\mathrm{P}$ nor $R_\mathrm{out}=7.66r_\mathrm{h}$ means anything for the Dehnen models. So we faced a problem in choosing some universal outer edge to measure the global SFE in order to parameterize our Dehnen model clusters. 
If we choose the same logic as before and try to catch at least 98 percent of stellar mass for the Dehnen models, the outer radius might go too much far out of cluster (e.g $>38.5r_\mathrm{h}$ for $\gamma=0$ or $>49r_\mathrm{h}$ for $\gamma=2$, see Fig.~\ref{fig:mass-prof}). 
Therefore we propose 2 options to measure SFE representing the cluster globally in the scope of this paper. The first, following \citet{Bek+17} we use the global SFE redefined as
\begin{equation}\label{eq:SFE_gl}
    \sfeD=\mathrm{SFE_r}(10a_\star),
\end{equation}
where $a_\star$ is the scale radius of the chosen stellar density profile. For the Dehnen models, mass fraction enclosed within $10a_\star\equiv10a_\mathrm{D}$ varies from 0.75 to 0.96 for different $\gamma$ (see Fig.~\ref{fig:massfrac}). 

\begin{figure}[ht!]
    \includegraphics[width=\columnwidth]{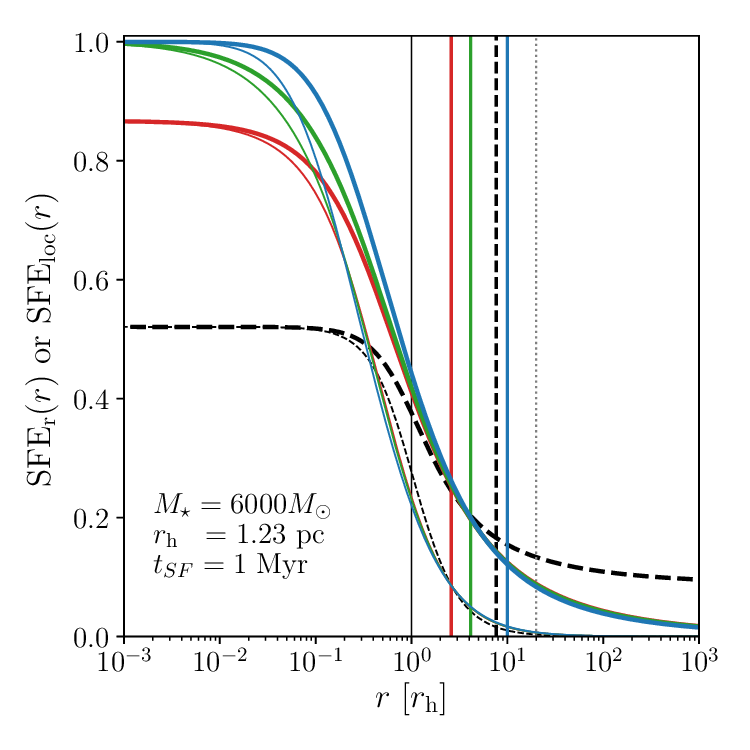}
    \caption{The cumulative and the local SFE profiles of model clusters with different stellar density profiles corresponding to  the Dehnen model ($\gamma=0,1,2$ in red, green and blue solid lines) and the Plummer model (dashed lines), given in thick and thin lines, respectively. SFE per free-fall time of $\epsilon_\mathrm{ff}=0.05$ and the SF duration of $t_\mathrm{SF}=1$~Myr (or equivalently $\epsilon_\mathrm{ff}=0.01$ and $t_\mathrm{SF}=5$~Myr) have been adopted to recover the density profiles of the residual gas for a cluster with $M_\star=6000\msol$ and $r_\mathrm{h}=1.23\pc$. The vertical lines correspond to the stellar half-mass radius, $r_\mathrm{h}$ (black solid line), $10 a_\mathrm{D}$ (red, green and blue solid lines corresponding to $\gamma=0,1$ and 2), $10a_\mathrm{P}$ (black dashed line), and the Jacobi radius, $R_\mathrm{J}$ (black dotted line), respectively. \label{fig:sfe-prof}}
\end{figure}

Another option is to measure the SFE within the Jacobi radius of the stellar cluster, $R_\mathrm{J}$, ignoring the residual gas mass. We understand that this definition of SFE as Jacobi SFE cannot be universal, since the Jacobi radius of the same star cluster can vary depending on the adopted impact of the Galactic tidal field. Since we aim to compare our new results with the default (or so-called standard) models of \citet{Bek+17} \citep[see also ][]{Bek+19} we adopt the impact of the tidal field of 
\begin{equation}
    \lambda = r_\mathrm{h}/R_\mathrm{J} = 0.05.
\end{equation}
{ The adopted value of $\lambda=0.05$ might seem to be too small compared to that of the observed open clusters varying around 0.1 and 0.2 in most cases \citep{Piskunov+2008, PZ+2010review, Fujii+PZ2016}. However, for the Solar neighborhood $\lambda=0.05$ results in the mean density within half-mass radius, $\rho_\mathrm{h}\approx 388\msol\pc^{-3}$, which is still in the range of the observed mean densities for clusters younger than 5 Myr \citep{LL03,PZ+2010review,Fujii+PZ2016,Krumholz+19}. According to \citet{MarksKroupa2012} the observed open clusters were much denser (about $10^4\msol\pc^{-3}$) at their gas embedded phases. Also the star formation simulations of \citet{Fujii+PZ2016,Sirius2,sirius3} results in higher densities ($>10^2\msol\pc^{-3}$) for embedded clusters than for gas-free clusters after gas expulsion. Therefore, the adopted value of $\lambda$ is consistent with other studies of embedded clusters.
}
In the scope of this study, we consider model star clusters with the same stellar mass, $M_\star$, and the stellar half-mass radius, $r_\mathrm{h}$. Thus the Jacobi SFE of our model clusters is 
\begin{equation}\label{eq:SFE_RJ}
 \mathrm{SFE_{J}}=\mathrm{SFE_r}(R_\mathrm{J})\equiv\mathrm{SFE_r}(20r_\mathrm{h}).   
\end{equation}
We have chosen the Jacobi SFE { to parameterize our Dehnen models}, because it accounts for more than 95 percent of the stellar mass (for $\gamma\leq 2.1$, Fig.~\ref{fig:massfrac}), the local SFE drops down to about 0.6 percent (Fig.~\ref{fig:sfe-prof}) at $R_\text{J}$, it does not depend on $\gamma$ and also it has a physical meaning for our model clusters. 
{ We plan to discuss more about the robust measurements of the SFE and the outer truncation radius in the follow-up paper. In this study, we consider some simplest ways of measuring the SFE.}
So in Fig. \ref{fig:sfe-prof}
we demonstrate the cumulative SFE (Eq.\ref{eq:SFE_r}) and local SFE (Eq.\ref{eq:SFE_loc}) profiles of model clusters distributed by the Plummer model (black dashed lines) and the Dehnen models ($\gamma=0,1,2$ in red, green and blue solid lines) in thick and thin lines, respectively. { Interestingly, Dehnen models with different $\gamma$ result in a very similar cumulative SFE at $r>r_\mathrm{h}$ if the radius is given in the units of $r_\mathrm{h}$. The local SFEs are similar even for about $r>r_\mathrm{h}/2$. Therefore for Dehnen models with different inner slopes, measuring the global SFE within the radius $r=xr_\mathrm{h}$ is preferable, then within $r=ya_\mathrm{D}$, where $x$ and $y$ are arbitrary numbers larger than the unity.}

The SFE profiles in Fig. \ref{fig:sfe-prof}
\begin{figure}
    \centering
    \includegraphics[width=\columnwidth]{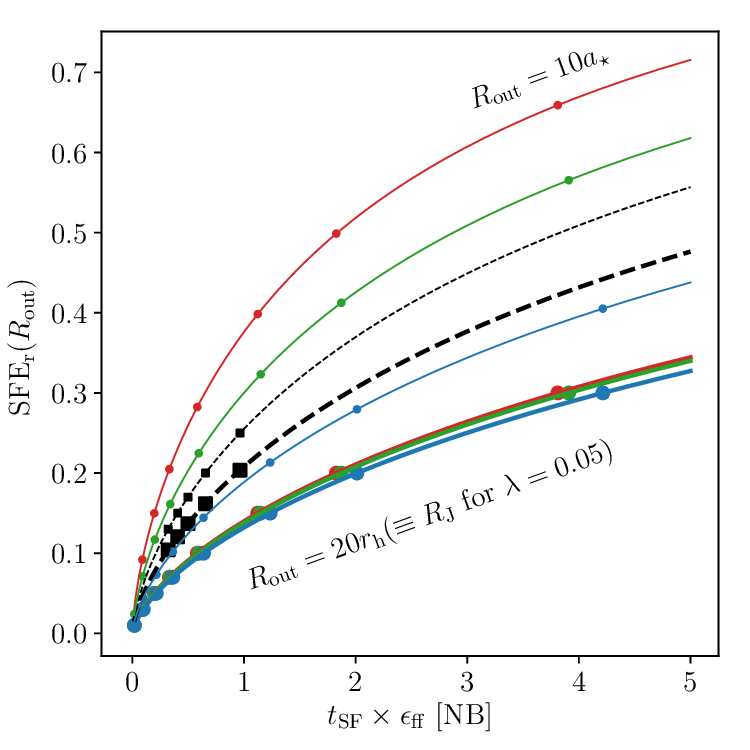}
    \caption{The ``global'' SFEs as measured by $\sfe$ and $\sfeD$ of model clusters presented as functions of $t_\mathrm{SF} \epsilon_\mathrm{ff}$, represented by red, green and blue solid lines for the Dehnen models with $\gamma=0,1,$ and 2, and dashed lines for the Plummer model, in thick and thin lines, respectively. The marked points correspond to the parameterization values of $\sfe$ for the Dehnen models, and $\sfeP$ for the Plummer models, respectively. }
    \label{fig:sfe-tsf}
\end{figure}
have been calculated for model clusters with mass $M_\star=6000\msol$, half-mass radius $r_\mathrm{h}=1.23$ for $\epsilon_\mathrm{ff}=0.05$ and $t_\mathrm{SF}=1$~Myr (or equivalently for $\epsilon_\mathrm{ff}=0.01$ and $t_\mathrm{SF}=5$~Myr). The chosen cluster half-mass radius corresponds to $\lambda=0.05$ when a cluster moves on circular orbit on the Galactic disk plane at the Galactocentric distance of the Sun, $R_\mathrm{orb}=8178\pc$ \citep[][]{GravityColl2019}. We calculate the Jacobi radius of our stellar clusters according to \citet[][Eq.13]{Just+09}
\begin{equation}\label{eq:RJ}
    R_\mathrm{J}=\left(\frac{GM_\star}{(4-\beta^2)\Omega^2}\right)^{1/3} \approx 24.52\pc.
\end{equation}
Here $\beta=1.37$ is the normalized epicyclic frequency, $\Omega=V_\mathrm{orb}/R_\mathrm{orb}$ is the angular speed of star cluster on a circular orbit \citep{Just+09} at a distance $R_\mathrm{orb}=8178\pc$ with the circular speed $V_\mathrm{orb}=234.73\ \mathrm{km\ s^{-1}}$. We assume here that the total stellar mass resides inside the Jacobi radius, although small fraction of stars might be outside depending on the density profile. 

{Figure \ref{fig:allSFEs} 
\begin{figure*}
    \centering
    \includegraphics[width=\linewidth]{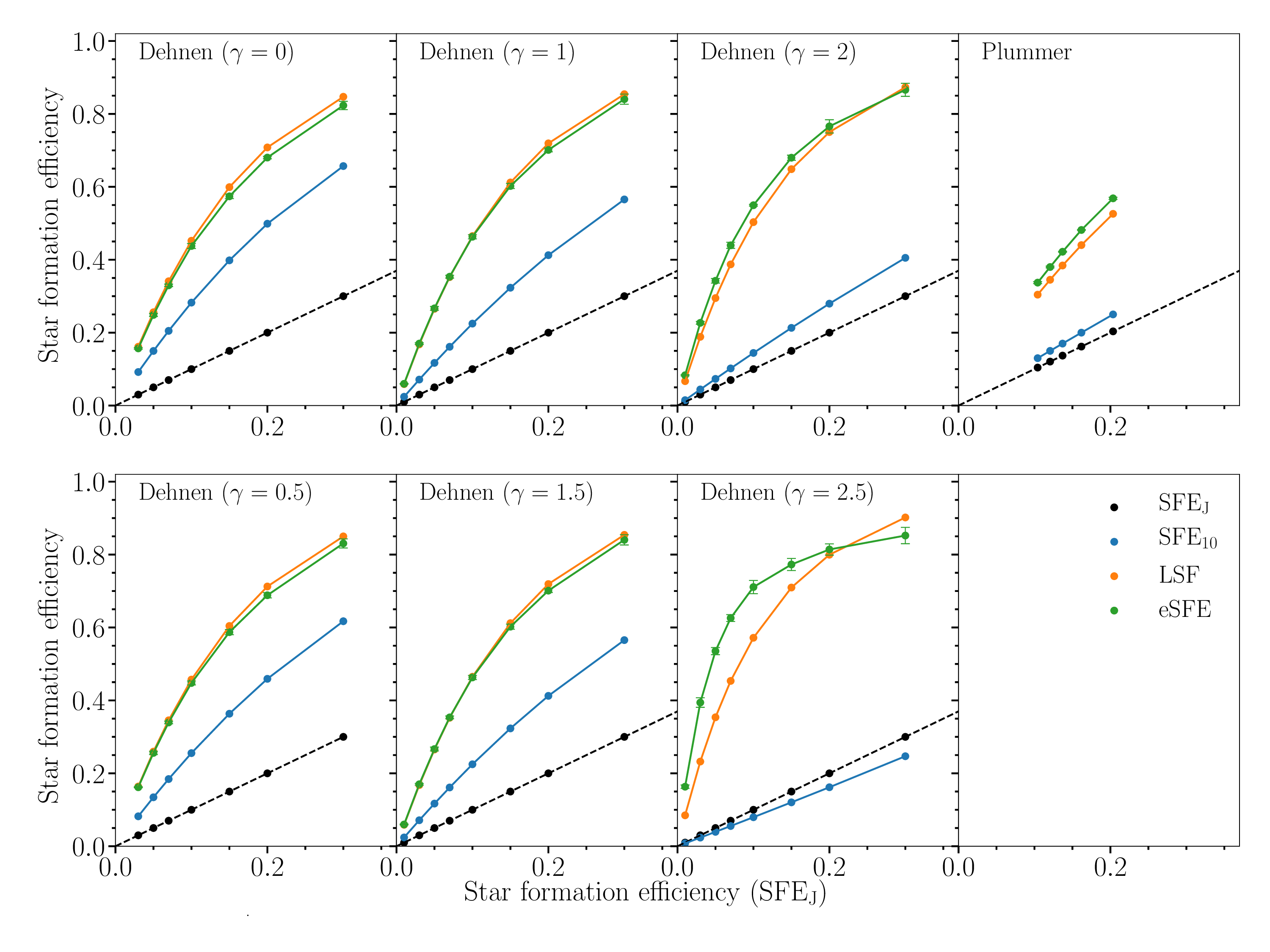}
    \caption{ Comparison of differently measured SFEs: $\sfe$, $\sfeD$, eSFE and LSF for the considered cluster models. The Jacobi SFE $(\mathrm{SFE_J})$ has been chosen for parameterization of differently measured SFEs and is plotted in black 1 to 1 line.}
    \label{fig:allSFEs}
\end{figure*}
presents the relation between differently measured SFEs ($\sfe$, $\sfeD$, eSFE and LSF) for the Plummer and the Dehnen models. We note here that the $\sfe$ is the only SFE which depends on the environment. Other SFEs are independent of the impact of the tidal field of the host galaxy, but depend on the cluster density profile. Therefore, when we refer to the $\sfe$, we mean the Jacobi SFE for $\lambda=0.05$ in future, i.e. $\sfe=\mathrm{SFE_r}(20r_\mathrm{h})$.}

\subsection{$N$-body simulations}


We use the high-precision $\phi$-GRAPE/GPU \citep{Harfst+07,Berczik+13} direct $N$-body code with 4-th order Hermite integrator, which uses GPU/CUDA based GRAPE emulation YEBISU library \citep{Nitadori+Makino08} to perform our $N$-body simulations. The binary stars are not considered in this code, instead the small enough softening parameter $\epsilon=10^{-4}$NB has been introduced to avoid close encounters of stars. The evolution of single stars{, including the mass-loss from the stellar wind} are accounted through the updated \textsc{sse} code \citep{Hurley+00,Banerjee+2020,Kamlah+2021} integrated to $\phi$-GRAPE/GPU code. In particular we use the level C \textsc{sse} code from \citet{Kamlah+2021}, which mainly differs from the original \textsc{sse} code of \citet{Hurley+00} on the stellar evolution prescriptions of massive stars. The new prescriptions include remnant-mass and fallback according to the rapid and delayed SNe models of \citet{Fryer+2012}, the pair-instability SNe \citep{Belczynski+2016} and Electron Capture SNe with small neutron star kicks \citep{Belczynski+2008}. For more details see \citet{Kamlah+2021}.
In contrast to previous simulations \citep{Bek+17,Bek+18,Bek+19}, we take into account the natal-kick velocities of the SNe remnants in this study. Due to these differences in stellar evolutionary prescriptions we decided to redo the simulations with the Plummer model clusters in the scope of this study.
{ For the Galactic potential in our simulations we use the three-component (bulge-disk-halo) axisymmetric Plummer-Kuzmin model \citep{MiyamotoNagai75} alike used in \citet{Just+09} and our previous works \citep[e.g. ][]{Bek+17,Bek+18,Bek+19}
\begin{equation}
  \Phi(R,z) = -\frac{GM}{\sqrt{R^2+\left(a+\sqrt{b^2+z^2}\right)^2}},
\end{equation}
where $M, a, b$ are the mass, flattening parameter and core radius of the component \citep[see Table 1 in ][]{Bek+19}. However, in this study we slightly tuned the halo mass to $M_\mathrm{halo}=7.2535\times 10^{11}\msol$ to get the circular speed $V_\mathrm{orb}=234.73 \mathrm{km\ s^{-1}}$ at the Galactocentric distance of the Sun, $R_\mathrm{orb}=8178\pc$ \citep[][]{GravityColl2019}. Other parameters are kept as in Table 1 of \citet{Bek+19}.}

We adopted the following $N$-body units (NBU) and constants in our simulations of Dehnen clusters:
\begin{equation}
    G=M_\star=r_\mathrm{h}=1\ [\mathrm{NBU}].
\end{equation}
Model clusters (either with the Dehnen profiles or the Plummer profile) in the scope of this study have $N_\star=10455$ stars, leading to $M_\star = 6000\msol$ when the stellar IMF \citep[][]{Kroupa2001} with upper and lower stellar masses of $m_\mathrm{up}=100\msol$ and $m_\mathrm{low}=0.08\msol$ is applied. The $N$-body units for the newly ran Plummer model simulations kept as in \citet{Bek+17}.

\subsection{Random realizations}

We consider simulations of star cluster models for different $\sfe=[0.01,0.03,0.05,0.07,0.1,0.15,0.2,0.3]$ and 6 different $\gamma=0,0.5,1,1.5,2,2.5$ with the Dehnen profiles and different $\sfeP=[0.13,0.15,0.17,0.2,0.25]$ with the Plummer profile. We did not consider models with very low $\sfe=0.01$ for the Dehnen profiles with $\gamma=0$ and $\gamma=0.5$. For each model, we obtain 9 random realizations, where we randomize the phase-space distribution of stars with 3 random seeds and the stellar IMF with 3 random seeds. That is, from \textsc{mkhalo} we get 3 different models of single-mass $N$-body systems for a given set of parameters (model, i.e. Plummer or Dehnen with a given $\gamma$, and $\sfe$ defined by the product of $\epsilon_\mathrm{ff}$ and $t_\mathrm{SF}$ ). Then for each model we assign stellar masses according to stellar IMF \citet{Kroupa2001}. In total, we have performed 459 simulations ($9\cdot(7\cdot 2+8\cdot 4)+5\cdot 9$), including 414 simulations of Dehnen models and 45 simulations of Plummer models. When presenting results, we demonstrate the mean and standard deviation of considered parameters (e.g. bound mass fraction) from the sample of 9 random realization of a given model.

\section{Results}\label{sec:res}

\begin{figure*}[h!]
    \centering
    \includegraphics[width=\hsize]{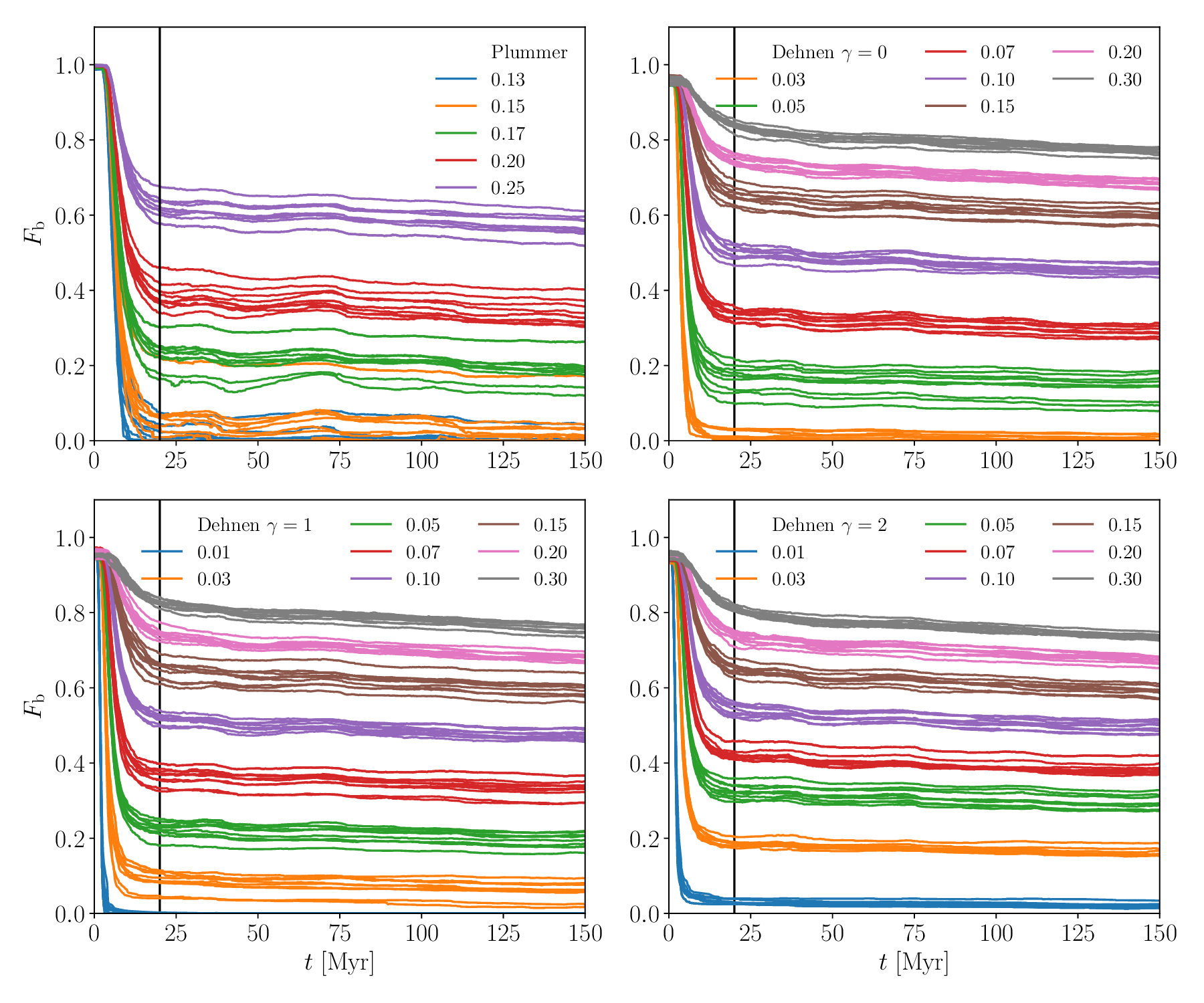}
    \caption{The bound mass fraction evolution of clusters corresponding to the Plummer model (upper left panel) and the Dehnen ($\gamma=0,1,2$ as indicated in legends) models during the first 150 Myr after instantaneous gas expulsion. The lines are color-coded by the corresponding SFEs as in the key. $\sfeP$ has been used for the Plummer clusters, while $\sfe$ for the Dehnen clusters, respectively. The adopted end of violent relaxation $t_\mathrm{VR}=20$~Myr is indicated by the vertical black line.}
    \label{fig:fbevol}
\end{figure*}

We have looked at the evolution of bound mass fractions, $F_\mathrm{b}$, of our star clusters within the first 150 Myr time-span as shown in Fig \ref{fig:fbevol}. 
As in our previous studies, we define the bound-mass fraction as a ratio of the Jacobi mass at the current time, $M_\mathrm{J}$, to the total initial stellar mass at the time of instantaneous gas expulsion, $M_\star$
\begin{equation}
    F_\mathrm{b}(t) = \frac{M_\mathrm{J}(t)}{M_\star}.
\end{equation}
The upper left panel of Fig. \ref{fig:fbevol} corresponds to the new Plummer (with SNe natal kick and updated \textsc{sse}) and the other panels to the Dehnen ($\gamma=0,1,2$) model simulations, respectively. As we can see, there is a scatter from model randomization in both cases. \citet{Bek+18} showed that the most massive stars might cause sub-structure formation during the early expansion phase of cluster evolution after gas expulsion. In some cases these most massive stars can help clusters to survive with higher bound fraction than the average for a given SFE, as well as to leave the cluster with a low bound fraction by escaping early \citep[see Fig.~1 in ][]{Bek+18}.  In \citet{Bek+19} we have calculated the end of violent relaxation based on comparison of star cluster total mass-loss and the stellar evolutionary mass-loss. The assumed end of violent relaxation coincides with the moment when cluster total mass loss is almost equal to that of the stellar evolution. For the default models of \citet{Bek+19} with $\lambda=0.05$ (i.e. also S0-models), the end of violent relaxation has been estimated to be $t_\mathrm{VR}=17.9\pm 2.3$~Myr. In this study, for the simplicity, we have chosen $t_\mathrm{VR}=20$~Myr, as in \citet{Bek+17}, which is consistent with estimation of \citet{Bek+19}. From Fig.~\ref{fig:fbevol} it is clear that the difference of few Myr in the adopted time of the end of violent relaxation is still applicable. This $t_\mathrm{VR}=20$~Myr is  indicated by the vertical black line in of Fig.~\ref{fig:fbevol}, to the right from which the bound mass fraction becomes almost constant. The shapes of the bound mass fraction evolution in both models are quite similar to those presented in previous studies \citep{Bek+17, Bek+19}. The wave-like shape corresponds to the vertical oscillation of the vertically escaped stars in the disk potential, bypassing the cluster Jacobi radius vertically. It is moderately small in case of the Dehnen models, where relatively compact bound cluster forms after gas expulsion, compared to the case of the Plummer model.

In case of low-SFE Dehnen model clusters, we can see that the violent relaxation seems to end earlier, when compared to the Plummer models. The Dehnen clusters have more compact and denser core than Plummer clusters. Also the Dehnen model clusters contain much more amount of gas in the outer shells than the Plummer clusters. Thus the unbound stars of low-SFE Dehnen clusters escape faster leaving the compact core. More studies about structural change of our model clusters will be discussed in the upcoming papers. Nevertheless, $\tvr=20\Myr$ seems to be quite good estimation of the end of violent relaxation, when the final bound mass fraction, $F_\mathrm{b20}$, is in focus. 

{}{ In Fig.~\ref{fig:Fb-SFE}, we present the final bound mass fraction of our model clusters (i.e. at the end of violent relaxation $t_\mathrm{VR}=20$~Myr) as a function of SFEs, measured in different ways.
\begin{figure*}[!h]
    \centering
    \includegraphics[width=\hsize]{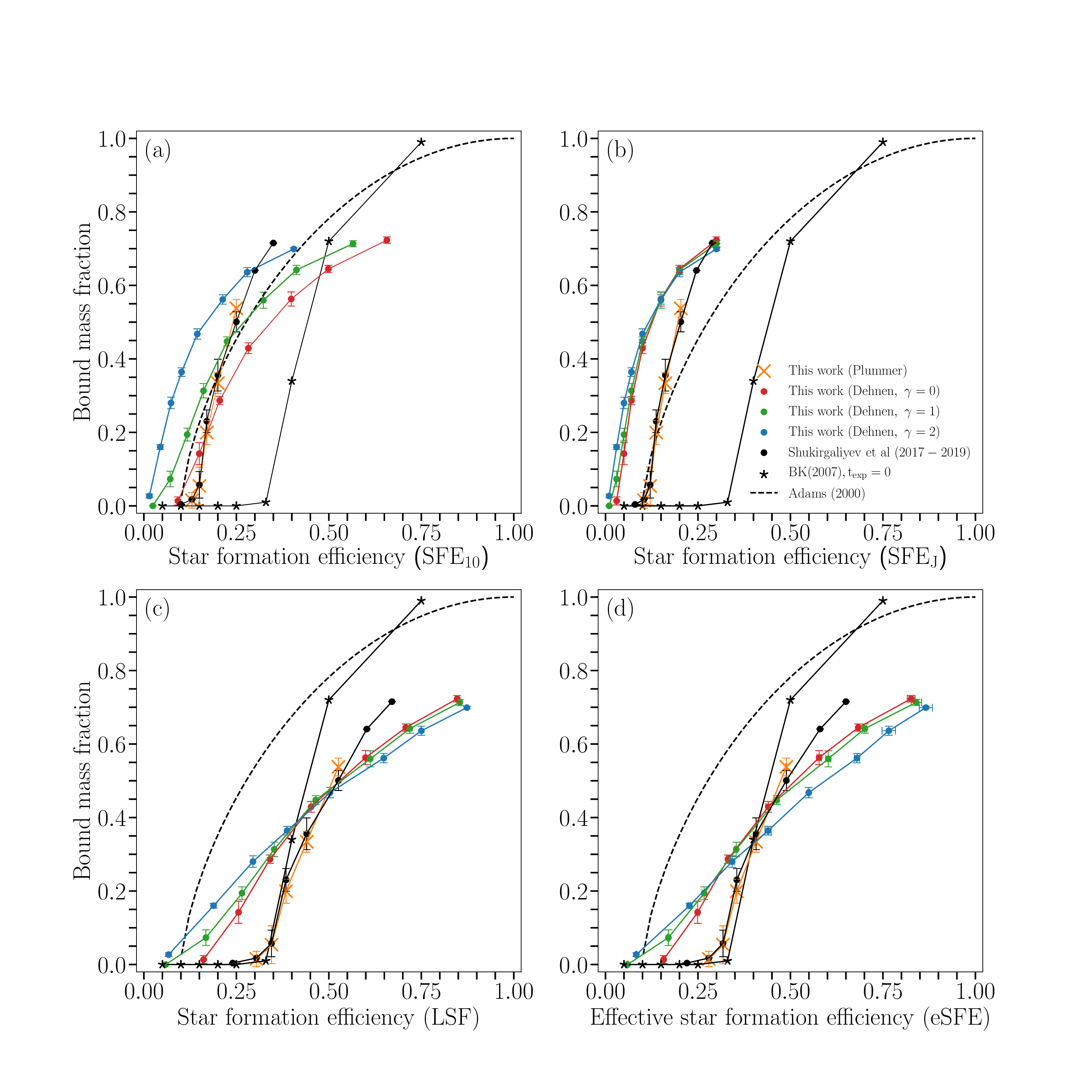}
    \caption{SFE and bound mass fraction at the end of violent relaxation $(t=\tvr)$ for the Dehnen models ($\gamma=0,1,2$ in red, green and blue points) and the Plummer model (orange crosses). The points correspond to the mean bound mass fraction of all random realizations of one model cluster, and error-bars correspond to the respective standard deviations. There are four different measurements of SFE considered in this plot. Panel (a) corresponds to SFE measured within 10 scale radii corresponding to the models, i.e. $\sfeP=\mathrm{SFE_r}(10a_\star)$ as it was  measured in \citet{Bek+17}. 
    Panel (b) presents SFE measured within the Jacobi radius, $\sfe = \mathrm{SFE_r}(R_\mathrm{J}\equiv20r_\mathrm{h})$. Panels (c) and (d) present the survivability of model clusters as a functions of LSF and eSFE. The black crosses correspond to the default model of \citet{Bek+17,Bek+19}, where the SNe remnant kick velocity is neglected. The results of \citet{Adams2000} (dashes black line) and \citet{BK07} (black star symbols, the instantaneous expulsion case only) are plotted for the reference as given in the original publications. For the case of \citet{BK07}, SFEs measured in all different ways are identical, because the density profiles of gas and stars have the same shapes.}
    \label{fig:Fb-SFE}
\end{figure*}
The survivability of model clusters are shown on the aspects of 4 different SFEs: on panel (a) of Fig~\ref{fig:Fb-SFE} SFE is measured within $10a_\mathrm{P}$ for the Plummer models and $10a_\mathrm{D}$ for the Dehnen models, in the panel (b) we measure SFE within a Jacobi radius, $<R_\mathrm{J}\equiv20r_\mathrm{h}$ for our simulations. In panels (c) and (d) we present the results in the aspect of LSF (i.e. SFE within a half-mass radius) and the eSFE (i.e. SFE based on virial ratio). The results for the Dehnen model clusters presented by red, green and blue dots corresponding to $\gamma=0,1,2$, and for newly ran Plummer models by blue plus symbols. The black crosses correspond to the default (or so-called ``standard'') models of the previous Plummer model clusters \citep{Bek+17,Bek+19}. Additionally in all panels results from \citet{Adams2000} (dashed line) and \citet{BK07} (black star symbols, their instantaneous gas expulsion case) have been presented as in the original papers, without any changes for the reference. In case of \citet{BK07} model, SFEs measured with different methods are equivalent to each other, because the stars and gas have density profiles of the same shape. Therefore it does not matter within which radius you measure SFE, it will stay constant globally or locally in the scope of a given model. For their models eSFE also equivalent to the total SFE \citep[see][]{GoodwinBastian2006, Goodwin2009}. Therefore, their SFEs are quite consistent with those presented in Fig~\ref{fig:Fb-SFE}.

We have found that our Plummer model simulations result in very similar final bound fraction, although the natal kick for SNe remnants has been included in the new simulations (compare orange crosses with black dots in Fig.~\ref{fig:Fb-SFE}). This is explained by the late occurrence of SNe (3 or more Myr after instantaneous gas expulsion), when the whole dynamics of the cluster is defined within the first few Myrs of expansion. Also, in the updated stellar evolution prescriptions high-mass SNe remnant black holes have increased masses and no natal kick due to fallback mechanism \citep{Belczynski+2016,Kamlah+2021}, and thus do not escape the cluster. Therefore, the difference in remaining bound mass fraction between old and new simulations is not significant. Nevertheless, the minimum SFE needed for the Plummer model clusters to survive instantaneous gas expulsion stays in 0.15, if measured within $10a_\mathrm{P}$, or lowered to 0.12 if measured within a Jacobi radius (only due to the measuring method of SFE). Corresponding values of differently measured SFEs can be found in the electronic table attached to the paper or on Fig.~\ref{fig:allSFEs}. In case of LSF and eSFE, the survival SFE threshold are about 0.34 and 0.32, respectively, for the Plummer model clusters. Thus they stay consistent with other simulations used Plummer model as initial condition for their star cluster simulations \citep[][]{GoodwinBastian2006,BK07,Smith+11,Brinkmann+17,LeeGoodwin2016}. 

The Dehnen model clusters show much better survivability than the Plummer model clusters (see red, green and blue dots in Fig.\ref{fig:Fb-SFE}) in all aspects of SFEs, especially for low-SFE cases. The minimum SFE required to survive the instantaneous gas expulsion is 0.03 or 0.09 for the Dehnen $\gamma=0$ model, depending on SFE measurement method, within $<R_\mathrm{J}$ or $<10a_\mathrm{D}$, respectively. For $\sfe=0.03\equiv\sfeP\approx0.05$ model some random realizations of cluster models could not survive after violent relaxation. However, model clusters with a bit higher SFE ($\sfe=0.05$) confidently survive the instantaneous gas expulsion with bound mass fraction of above 0.10.
{ The results from Dehnen models are quite similar when we look on the aspect of $\sfe$, but very different when the global SFE is represented by $\sfeD$. This difference occurs only due to the method of the SFE measurement. As we noticed in Fig.~\ref{fig:den-prof}, Dehnen models show very similar SFE values when the cumulative SFE is measured within the multiple of the half-mass radius.  }
The Dehnen models show good ability to survive the gas expulsion with low SFEs in the aspect of LSF and eSFE (see lower panels of Fig \ref{fig:Fb-SFE}). In this sense, results obtained from the Dehnen models in the aspect of LSF also resembles the results obtained from hierarchically formed highly sub-structured cluster models of \citet{Farias+18} with live gaseous background or results of \citet{Li+2019} obtained from their hydro-dynamical simulations of star cluster formation followed by $N$-body simulations. Although our models are calculated for the instantaneous gas expulsion, while other two studies are done for more realistic (stellar feedback-driven) 
gradual gas expulsion cases. 

In cases when SFEs are measured within 10 model scale radii, the results on survivability of the Dehnen $(\gamma=0)$ model and the Plummer model clusters are somewhat similar to each other in lower values. But high-SFE clusters of the Dehnen models then show lower bound fractions than the Plummer model clusters. 

Increasing the slope of the inner power law profile in Dehnen models also helps to survive the consequences of the instantaneous gas expulsion with very low global SFE (i.e. $\sfe$). Such an effect is already expected from their SFE profiles (see Fig.~\ref{fig:sfe-prof}), i.e. the Dehnen models with higher $\gamma$ have steeper SFE-profiles. Although, varying the inner power law slope does not make very big difference for higher-SFE model clusters (see Fig.~\ref{fig:Fb-gamma}).

In Fig.~\ref{fig:Fb-gamma}, we also present the differences between the final bound mass fractions of the Dehnen models across $\gamma$ values. 
\begin{figure}
    \centering
    \includegraphics[width=\columnwidth]{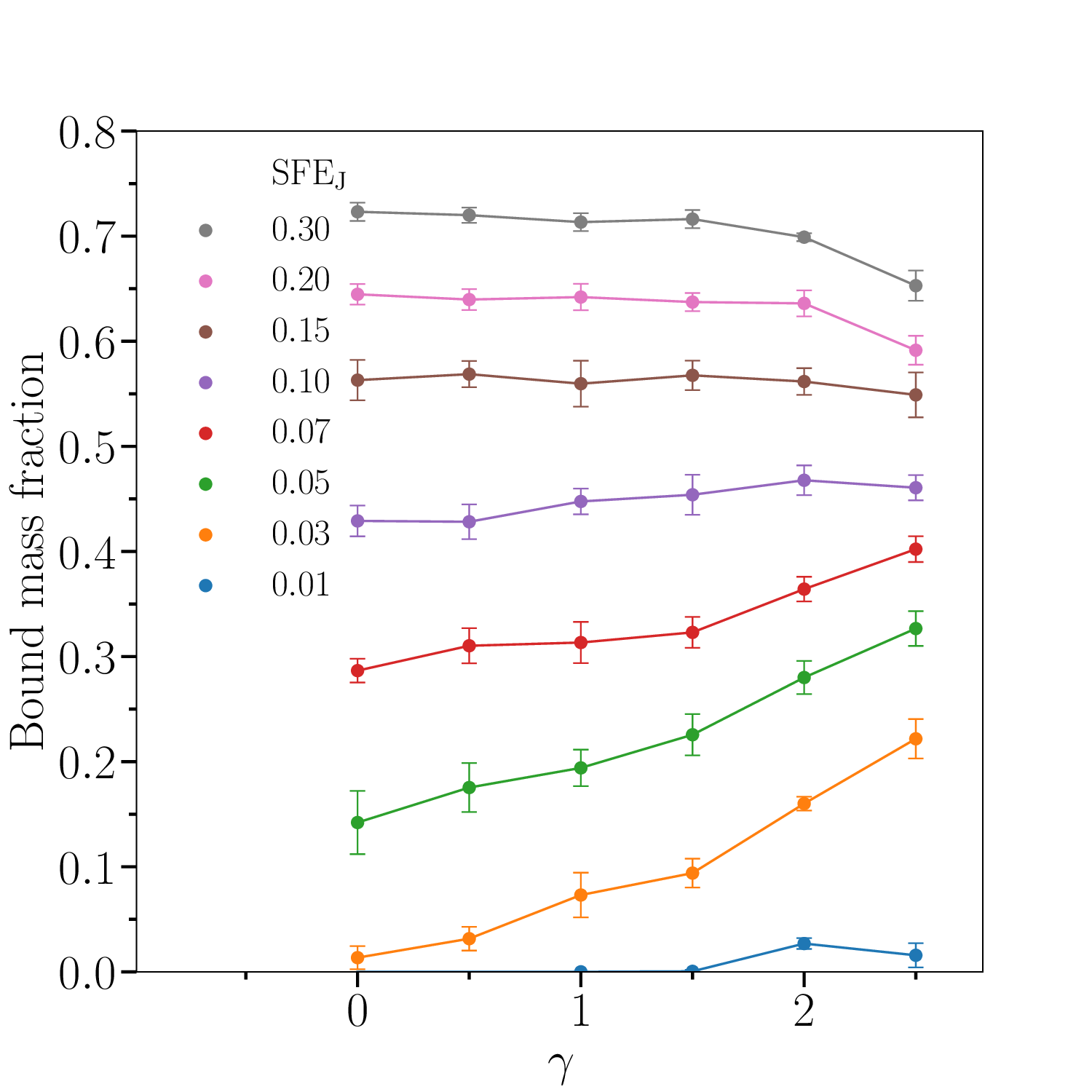}
    \caption{Comparison of the bound mass fractions at the end of violent relaxation $(t=\tvr)$ of the Dehnen model clusters with different $\gamma$. Models with the same $\sfe$ are indicated with corresponding colors as in the key.}
    \label{fig:Fb-gamma}
\end{figure}
It shows nice increasing trends of bound fraction with gamma for low-SFE clusters ($\sfe<0.15$). This indicates that having a cusp at the moment of formation can help these low-SFE clusters to survive the consequences of gas expulsion quite well. The decreasing of the final bound fraction at higher $\gamma$ for high-SFE clusters happens due to the fact that their scale radii are larger than their half-mass radii. Thus, a significant fraction of star becomes unbound quickly after gas expulsion because of their large distance from the cluster center. For $\gamma=2.5$, already 6 percent of the stellar mass becomes unbound immediately after gas expulsion staying beyond the new Jacobi radius (see Fig.~\ref{fig:massfrac}). 
}

We think that better survivability of the Dehnen model clusters, compared to the Plummer model ones, caused by the relative shallowness of the density profiles of both, gas and stars. The latter results in more centrally peaked (steeper) SFE-profile in case of the Dehnen model clusters (see Fig. \ref{fig:sfe-prof}).
\citet{Bek+19} already have shown that varying the cluster density, thus the impact of the Galactic tidal field does not impact significantly on the survivability of star clusters after gas expulsion. Therefore we think that the steepness of the slope of SFE profile plays a significant role on the survivability of the stellar clusters after gas expulsion and subsequent violent relaxation, rather than the cluster central density.

\section{Conclusions and Discussions}\label{sec:conc}

In this study we consider the survivability of star clusters with different density profiles after instantaneous gas expulsion. Namely we consider model clusters corresponding to the Plummer model and the Dehnen model with $\gamma$ varying from 0 to 2.5 in virial equilibrium within the residual gas immediately before instantaneous gas expulsion. The density profiles of the residual star-forming gas corresponding to a given global SFE have been recovered assuming that stars formed with a constant efficiency per free-fall time \citep{PP13,Bek+17}. Then we perform direct $N$-body simulations of clusters evolution after instantaneous gas expulsion, as if they are orbiting in the solar orbit, but exactly in the equatorial Galactic disk plane.   

We re-simulate the Plummer model clusters with the updated stellar evolution prescriptions \citep{Banerjee+2020,Kamlah+2021} and this time the natal kick of SNe remnants is included. As we see from the results of our simulations, the accounting for or neglecting the natal kick velocity of SNe remnants do not affect much the survivability of star clusters after gas expulsion. This is because SNe happen 3 Myr after the instantaneous gas expulsion or later, since we assume that all stars in our model cluster enters the main sequence exactly at the time of gas expulsion. { In real clusters there is an age spread of less than 5 Myr \citep{Reggiani+2011,Kudryavtseva+2012}. However, only a few of the most massive (O-B) stars entering the main sequence can start the gas expulsion. } The cluster dynamics after gas expulsion is set already by the gas potential. Then SNe happens in the already expanding cluster or already after violent relaxation when cluster is back to virial equilibrium with less amount of original number of stars. During the violent relaxation, only a part of massive stars undergo SNe.
Also, in the updated stellar evolution prescriptions high-mass black hole remnants have zero kick velocity and increased mass due to the fallback mechanism \citep{Belczynski+2008,Belczynski+2016,Banerjee+2020,Kamlah+2021}.
Therefore SNe natal kick does not impact the final bound mass fraction, however can be significant in cluster structure and affect the long-term evolution, which are not considered in the scope of this study.

In this study, we concentrate our attention on the question, whether the Dehnen model clusters survive the gas expulsion in the same way as previously considered Plummer model clusters \citep{Bek+17,Bek+19} or differently. Also we discuss in this study the problem of measuring the global SFE, both in theory and observations.
The problem emerges when density profiles of stars and residual gas do not follow the same shape. In nearby star-forming regions \citet{Gutermuth+2011}, and recently \citet{Pokhrel+20} found  the power law correlation between surface densities of stars and gas $\Sigma_\star\propto\Sigma_{gas}^2$. That is stellar density profile has a steeper slope than that of gas. \citet{PP13} explains such a behavior of gas and stars by their local density-driven clustered star formation model, where star-formation happens with a constant efficiency per free-fall time. As a consequence, in a given time-span dense gas produces more stars, than the diffuse gas. In such systems the local SFE is not equivalent to the global SFE, as it was in models of \citet{BK07} and references therein. Also, the global SFE depends on the radius of sphere of measurement, decreases with increasing radius. Two different SFE measurements representing the global SFE are considered: one measured within stellar cluster Jacobi radius, $R_\mathrm{J}$, ignoring the gas mass ($\sfe$), and another measured as in \citet{Bek+17}, within 10 times model scale radii, $\sfeP = \mathrm{SFE_r}(10a_\star)$. 
Also we consider the local stellar fraction (LSF) -- SFE measured within the stellar half-mass radius, $r_\mathrm{h}$, and the effective SFE (eSFE) -- SFE defined based on the cluster virial state immediately after gas expulsion (see Eqs. \ref{eq:LSF}-\ref{eq:eSFE} and the text in sec.~\ref{sec:intro} for details).
{ The Jacobi SFE is considered for the case when $\lambda=r_\mathrm{h}/R_\mathrm{J}=0.05$, thus $\sfe=\mathrm{SFE_r}(20r_\mathrm{h})$. That is the Jacobi SFE is the only method, which is sensitive on the adopted environment (i.e. on $\lambda$). We also noticed, that the Dehnen models with different inner slopes, $\gamma$, result in similar values of cumulative SFE, if measured within the multiple of a half-mass radius (see Fig.~\ref{fig:den-prof}).}

In sec.~\ref{sec:res} we show the bound mass fraction of our model clusters, measured at the end of violent relaxation ($t=20$~Myr) as functions of considered SFEs (see Fig.~\ref{fig:Fb-SFE}). We have found that the Dehnen model clusters have better ability to survive gas expulsion with very low SFE than the Plummer model clusters. {}{ The minimum SFE needed to survive can be as low as $\sfe=0.01$ for the Dehnen model clusters with $\gamma=2\mathrm{\ and\ }2.5$. }This is because the Dehnen model clusters have higher SFE in the inner part and lower SFE in the outer part when compared to the Plummer model. That is a consequence of having shallower density profiles for gas and stars in case of the Dehnen models than in case of Plummer models. Hence the shallower the slope of stellar density profile, the lower the critical global SFE needed to survive the instantaneous gas expulsion. This statement applies to gas embedded stellar clusters formed in centrally-concentrated spherically-symmetric gas clumps with a constant efficiency per free-fall time. 

We conclude that the shallowness of the outer power law density profile of stellar cluster helps it to survive the instantaneous gas expulsion with low SFEs. Also star clusters with a cusp survive the consequences of instantaneous gas expulsion better than those without. Additionally, the higher the slope of the cusp density power law profile, the higher the fraction of stellar mass remains bound to the cluster after violent relaxation.

\begin{acknowledgements}
We thank the referee for the helpful comments.

This research has been funded by the Science Committee of the Ministry of Education and Science of the Republic of Kazakhstan (Grant No. AP08856149, AP08052197, and AP08856184).

This work was supported by the Deutsche Forschungsgemeinschaft (DFG, German Research Foundation) -- Project-ID 138713538 -- SFB 881 (‘The Milky Way System’) sub-project B2 -- through the individual research grant 'The dynamics of stellar-mass black holes in dense stellar systems and their role in gravitational-wave generation' (BA 4281/6-1).

This work was partly supported by the International Partnership Program of Chinese Academy of Sciences, Grant No. 114A11KYSB20170015 (GJHZ1810). BS, MS, TP, and MK are grateful for hospitality and support during visits to Silk Road Project of National Astronomical Observatories of Chinese Academy of Sciences.

MS, MI, EP, PB, RS and AJ acknowledge the partial support by the Volkswagen Foundation under the Trilateral Partnerships grant No. 97778. 

PB and MI acknowledges the support by the National Academy of Sciences of Ukraine under the "Application of the high-performance parallel cluster computing in astrophysics and space geodesy", No. 13.2021.MM.

OB and EP acknowledges joint RFBR and DFG research project 20-52-12009 (N-body simulations).

TP acknowledges the support by the European Research Council (ERC) under the European Union’s Horizon 2020 research and innovation program under grant agreement No 638435 (GalNUC).

PB acknowledges support by the Chinese Academy of Sciences through the Silk Road Project at NAOC, the President’s International Fellowship (PIFI) for Visiting Scientists program of CAS, the National Science Foundation of China under grant No. 11673032.

\\

Facilities: Super-computing facilities at the Energetic Cosmos Laboratory of Nazarbayev University, Fesenkov Astrophysical Institute, Main Astronomical Observatory of the National Academy of Sciences of Ukraine,  Zentrum f\"ur Astronomie der Universit\"at Heidelberg, INASAN and Silkroad project at the National Astronomical Observatories of China/Chinese Academy of Sciences have been used for calculations in this study.
\end{acknowledgements}

%
%

\bibliographystyle{aa}
\bibliography{aanda.bib}

\begin{thebibliography}{71}
\expandafter\ifx\csname natexlab\endcsname\relax\def\natexlab#1{#1}\fi

\bibitem[{{Adams}(2000)}]{Adams2000}
{Adams}, F.~C. 2000, \apj, 542, 964

\bibitem[{{Banerjee} {et~al.}(2020){Banerjee}, {Belczynski}, {Fryer},
  {Berczik}, {Hurley}, {Spurzem}, \& {Wang}}]{Banerjee+2020}
{Banerjee}, S., {Belczynski}, K., {Fryer}, C.~L., {et~al.} 2020, \aap, 639, A41

\bibitem[{{Baumgardt} \& {Kroupa}(2007)}]{BK07}
{Baumgardt}, H. \& {Kroupa}, P. 2007, \mnras, 380, 1589

\bibitem[{{Belczynski} {et~al.}(2016){Belczynski}, {Heger}, {Gladysz},
  {Ruiter}, {Woosley}, {Wiktorowicz}, {Chen}, {Bulik}, {O'Shaughnessy}, {Holz},
  {Fryer}, \& {Berti}}]{Belczynski+2016}
{Belczynski}, K., {Heger}, A., {Gladysz}, W., {et~al.} 2016, \aap, 594, A97

\bibitem[{{Belczynski} {et~al.}(2008){Belczynski}, {Kalogera}, {Rasio}, {Taam},
  {Zezas}, {Bulik}, {Maccarone}, \& {Ivanova}}]{Belczynski+2008}
{Belczynski}, K., {Kalogera}, V., {Rasio}, F.~A., {et~al.} 2008, \apjs, 174,
  223

\bibitem[{{Berczik} {et~al.}(2013){Berczik}, {Spurzem}, {Wang}, {Zhong}, \&
  {Huang}}]{Berczik+13}
{Berczik}, P., {Spurzem}, R., {Wang}, L., {Zhong}, S., \& {Huang}, S. 2013, in
  Third International Conference ''High Performance Computing, 52--59

\bibitem[{{Brinkmann} {et~al.}(2017){Brinkmann}, {Banerjee}, {Motwani}, \&
  {Kroupa}}]{Brinkmann+17}
{Brinkmann}, N., {Banerjee}, S., {Motwani}, B., \& {Kroupa}, P. 2017, \aap,
  600, A49

\bibitem[{{Chen} {et~al.}(2021){Chen}, {Li}, \& {Vogelsberger}}]{Chen+2021}
{Chen}, Y., {Li}, H., \& {Vogelsberger}, M. 2021, \mnras, 502, 6157

\bibitem[{{Dehnen}(1993)}]{dehnen_family_1993}
{Dehnen}, W. 1993, \mnras, 265, 250

\bibitem[{{Dehnen}(2000)}]{Dehnen2000}
{Dehnen}, W. 2000, \apjl, 536, L39

\bibitem[{{Dehnen}(2002)}]{Dehnen2002}
{Dehnen}, W. 2002, Journal of Computational Physics, 179, 27

\bibitem[{{Farias} {et~al.}(2018){Farias}, {Fellhauer}, {Smith},
  {Dom{\'\i}nguez}, \& {Dabringhausen}}]{Farias+18}
{Farias}, J.~P., {Fellhauer}, M., {Smith}, R., {Dom{\'\i}nguez}, R., \&
  {Dabringhausen}, J. 2018, \mnras, 476, 5341

\bibitem[{{Fryer} {et~al.}(2012){Fryer}, {Belczynski}, {Wiktorowicz},
  {Dominik}, {Kalogera}, \& {Holz}}]{Fryer+2012}
{Fryer}, C.~L., {Belczynski}, K., {Wiktorowicz}, G., {et~al.} 2012, \apj, 749,
  91

\bibitem[{{Fujii} \& {Portegies Zwart}(2016)}]{Fujii+PZ2016}
{Fujii}, M.~S. \& {Portegies Zwart}, S. 2016, \apj, 817, 4

\bibitem[{{Fujii} {et~al.}(2021{\natexlab{a}}){Fujii}, {Saitoh}, {Hirai}, \&
  {Wang}}]{sirius3}
{Fujii}, M.~S., {Saitoh}, T.~R., {Hirai}, Y., \& {Wang}, L. 2021{\natexlab{a}},
  \pasj [\eprint[arXiv]{2103.02829}]

\bibitem[{{Fujii} {et~al.}(2021{\natexlab{b}}){Fujii}, {Saitoh}, {Wang}, \&
  {Hirai}}]{Sirius2}
{Fujii}, M.~S., {Saitoh}, T.~R., {Wang}, L., \& {Hirai}, Y. 2021{\natexlab{b}},
  \pasj [\eprint[arXiv]{2101.05934}]

\bibitem[{{Fukushima} \& {Yajima}(2021)}]{Fukushima+Yajima2021}
{Fukushima}, H. \& {Yajima}, H. 2021, arXiv e-prints, arXiv:2104.10892

\bibitem[{{Geyer} \& {Burkert}(2001)}]{GeyerBurkert2001}
{Geyer}, M.~P. \& {Burkert}, A. 2001, \mnras, 323, 988

\bibitem[{{Goodwin}(2009)}]{Goodwin2009}
{Goodwin}, S.~P. 2009, \apss, 324, 259

\bibitem[{{Goodwin} \& {Bastian}(2006)}]{GoodwinBastian2006}
{Goodwin}, S.~P. \& {Bastian}, N. 2006, \mnras, 373, 752

\bibitem[{{Goodwin} \& {Whitworth}(2004)}]{GoodwinWhitworth2004}
{Goodwin}, S.~P. \& {Whitworth}, A.~P. 2004, \aap, 413, 929

\bibitem[{{Grasha} {et~al.}(2019){Grasha}, {Calzetti}, {Adamo}, {Kennicutt},
  {Elmegreen}, {Messa}, {Dale}, {Fedorenko}, {Mahadevan}, {Grebel},
  {Fumagalli}, {Kim}, {Dobbs}, {Gouliermis}, {Ashworth}, {Gallagher}, {Smith},
  {Tosi}, {Whitmore}, {Schinnerer}, {Colombo}, {Hughes}, {Leroy}, \&
  {Meidt}}]{Grasha+19}
{Grasha}, K., {Calzetti}, D., {Adamo}, A., {et~al.} 2019, \mnras, 483, 4707

\bibitem[{{Gravity Collaboration} {et~al.}(2019){Gravity Collaboration},
  {Abuter}, {Amorim}, {Baub{\"o}ck}, {Berger}, {Bonnet}, {Brandner},
  {Cl{\'e}net}, {Coud{\'e} Du Foresto}, {de Zeeuw}, {Dexter}, {Duvert},
  {Eckart}, {Eisenhauer}, {F{\"o}rster Schreiber}, {Garcia}, {Gao}, {Gendron},
  {Genzel}, {Gerhard}, {Gillessen}, {Habibi}, {Haubois}, {Henning}, {Hippler},
  {Horrobin}, {Jim{\'e}nez-Rosales}, {Jocou}, {Kervella}, {Lacour},
  {Lapeyr{\`e}re}, {Le Bouquin}, {L{\'e}na}, {Ott}, {Paumard}, {Perraut},
  {Perrin}, {Pfuhl}, {Rabien}, {Rodriguez Coira}, {Rousset}, {Scheithauer},
  {Sternberg}, {Straub}, {Straubmeier}, {Sturm}, {Tacconi}, {Vincent}, {von
  Fellenberg}, {Waisberg}, {Widmann}, {Wieprecht}, {Wiezorrek}, {Woillez}, \&
  {Yazici}}]{GravityColl2019}
{Gravity Collaboration}, {Abuter}, R., {Amorim}, A., {et~al.} 2019, \aap, 625,
  L10

\bibitem[{{Grudi{\'c}} {et~al.}(2021){Grudi{\'c}}, {Guszejnov}, {Hopkins},
  {Offner}, \& {Faucher-Gigu{\'e}re}}]{starforge}
{Grudi{\'c}}, M.~Y., {Guszejnov}, D., {Hopkins}, P.~F., {Offner}, S. S.~R., \&
  {Faucher-Gigu{\'e}re}, C.-A. 2021, \mnras [\eprint[arXiv]{2010.11254}]

\bibitem[{{Gutermuth} {et~al.}(2011){Gutermuth}, {Pipher}, {Megeath}, {Myers},
  {Allen}, \& {Allen}}]{Gutermuth+2011}
{Gutermuth}, R.~A., {Pipher}, J.~L., {Megeath}, S.~T., {et~al.} 2011, \apj,
  739, 84

\bibitem[{{Harfst} {et~al.}(2007){Harfst}, {Gualandris}, {Merritt}, {Spurzem},
  {Portegies Zwart}, \& {Berczik}}]{Harfst+07}
{Harfst}, S., {Gualandris}, A., {Merritt}, D., {et~al.} 2007, \na, 12, 357

\bibitem[{{Higuchi} {et~al.}(2009){Higuchi}, {Kurono}, {Saito}, \&
  {Kawabe}}]{Higuchi2009}
{Higuchi}, A.~E., {Kurono}, Y., {Saito}, M., \& {Kawabe}, R. 2009, \apj, 705,
  468

\bibitem[{{Hills}(1980)}]{Hills1980}
{Hills}, J.~G. 1980, \apj, 235, 986

\bibitem[{{Hurley} {et~al.}(2000){Hurley}, {Pols}, \& {Tout}}]{Hurley+00}
{Hurley}, J.~R., {Pols}, O.~R., \& {Tout}, C.~A. 2000, \mnras, 315, 543

\bibitem[{{Just} {et~al.}(2009){Just}, {Berczik}, {Petrov}, \&
  {Ernst}}]{Just+09}
{Just}, A., {Berczik}, P., {Petrov}, M.~I., \& {Ernst}, A. 2009, \mnras, 392,
  969

\bibitem[{{Kainulainen} {et~al.}(2014){Kainulainen}, {Federrath}, \&
  {Henning}}]{Kainulainen+14}
{Kainulainen}, J., {Federrath}, C., \& {Henning}, T. 2014, Science, 344, 183

\bibitem[{{Kamlah} {et~al.}(2021){Kamlah}, {Leveque}, {Spurzem}, {Arca Sedda},
  {Askar}, {Banerjee}, {Berczik}, {Giersz}, {Hurley}, {Belloni},
  {K{\"u}hmichel}, \& {Wang}}]{Kamlah+2021}
{Kamlah}, A.~W.~H., {Leveque}, A., {Spurzem}, R., {et~al.} 2021, arXiv
  e-prints, arXiv:2105.08067

\bibitem[{{Krause} {et~al.}(2020){Krause}, {Offner}, {Charbonnel}, {Gieles},
  {Klessen}, {V{\'a}zquez-Semadeni}, {Ballesteros-Paredes}, {Girichidis},
  {Kruijssen}, {Ward}, \& {Zinnecker}}]{Krause+20}
{Krause}, M. G.~H., {Offner}, S. S.~R., {Charbonnel}, C., {et~al.} 2020, \ssr,
  216, 64

\bibitem[{{Kroupa}(2001)}]{Kroupa2001}
{Kroupa}, P. 2001, \mnras, 322, 231

\bibitem[{{Kruijssen} {et~al.}(2019){Kruijssen}, {Schruba}, {Chevance},
  {Longmore}, {Hygate}, {Haydon}, {McLeod}, {Dalcanton}, {Tacconi}, \& {van
  Dishoeck}}]{Kruijssen+19}
{Kruijssen}, J.~M.~D., {Schruba}, A., {Chevance}, M., {et~al.} 2019, \nat, 569,
  519

\bibitem[{{Krumholz} \& {Matzner}(2009)}]{Krumholz+Matzner2009}
{Krumholz}, M.~R. \& {Matzner}, C.~D. 2009, \apj, 703, 1352

\bibitem[{{Krumholz} {et~al.}(2019){Krumholz}, {McKee}, \&
  {Bland-Hawthorn}}]{Krumholz+19}
{Krumholz}, M.~R., {McKee}, C.~F., \& {Bland-Hawthorn}, J. 2019, \araa, 57, 227

\bibitem[{{Kudryavtseva} {et~al.}(2012){Kudryavtseva}, {Brandner}, {Gennaro},
  {Rochau}, {Stolte}, {Andersen}, {Da Rio}, {Henning}, {Tognelli}, {Hogg},
  {Clark}, \& {Waters}}]{Kudryavtseva+2012}
{Kudryavtseva}, N., {Brandner}, W., {Gennaro}, M., {et~al.} 2012, \apjl, 750,
  L44

\bibitem[{{Lada} \& {Lada}(2003)}]{LL03}
{Lada}, C.~J. \& {Lada}, E.~A. 2003, Annual Review of Astronomy and
  Astrophysics, 41, 57

\bibitem[{{Lada} {et~al.}(2010){Lada}, {Lombardi}, \& {Alves}}]{Lada+2010}
{Lada}, C.~J., {Lombardi}, M., \& {Alves}, J.~F. 2010, \apj, 724, 687

\bibitem[{{Lada} {et~al.}(1984){Lada}, {Margulis}, \& {Dearborn}}]{Lada+1984}
{Lada}, C.~J., {Margulis}, M., \& {Dearborn}, D. 1984, \apj, 285, 141

\bibitem[{{Lee} \& {Goodwin}(2016)}]{LeeGoodwin2016}
{Lee}, P.~L. \& {Goodwin}, S.~P. 2016, \mnras, 460, 2997

\bibitem[{{Leisawitz} {et~al.}(1989){Leisawitz}, {Bash}, \&
  {Thaddeus}}]{Leisawitz1989}
{Leisawitz}, D., {Bash}, F.~N., \& {Thaddeus}, P. 1989, \apjs, 70, 731

\bibitem[{{Li} {et~al.}(2019){Li}, {Vogelsberger}, {Marinacci}, \&
  {Gnedin}}]{Li+2019}
{Li}, H., {Vogelsberger}, M., {Marinacci}, F., \& {Gnedin}, O.~Y. 2019, \mnras,
  487, 364

\bibitem[{{Marks} \& {Kroupa}(2012)}]{MarksKroupa2012}
{Marks}, M. \& {Kroupa}, P. 2012, \aap, 543, A8

\bibitem[{{McMillan} \& {Dehnen}(2007)}]{McmillanDehnen07}
{McMillan}, P.~J. \& {Dehnen}, W. 2007, \mnras, 378, 541

\bibitem[{{Miyamoto} \& {Nagai}(1975)}]{MiyamotoNagai75}
{Miyamoto}, M. \& {Nagai}, R. 1975, \pasj, 27, 533

\bibitem[{{Murray}(2011)}]{Murray2011}
{Murray}, N. 2011, \apj, 729, 133

\bibitem[{{Nitadori} \& {Makino}(2008)}]{Nitadori+Makino08}
{Nitadori}, K. \& {Makino}, J. 2008, \na, 13, 498

\bibitem[{{Parmentier} \& {Pasquali}(2020)}]{PP2020}
{Parmentier}, G. \& {Pasquali}, A. 2020, \apj, 903, 56

\bibitem[{{Parmentier} \& {Pfalzner}(2013)}]{PP13}
{Parmentier}, G. \& {Pfalzner}, S. 2013, \aap, 549, A132

\bibitem[{{Piskunov} {et~al.}(2008){Piskunov}, {Schilbach}, {Kharchenko},
  {R{\"o}ser}, \& {Scholz}}]{Piskunov+2008}
{Piskunov}, A.~E., {Schilbach}, E., {Kharchenko}, N.~V., {R{\"o}ser}, S., \&
  {Scholz}, R.~D. 2008, \aap, 477, 165

\bibitem[{{Plummer}(1911)}]{Plummer_1911}
{Plummer}, H.~C. 1911, \mnras, 71, 460

\bibitem[{{Pokhrel} {et~al.}(2020){Pokhrel}, {Gutermuth}, {Betti}, {Offner},
  {Myers}, {Megeath}, {Sokol}, {Ali}, {Allen}, {Allen}, {Dunham}, {Fischer},
  {Henning}, {Heyer}, {Hora}, {Pipher}, {Tobin}, \& {Wolk}}]{Pokhrel+20}
{Pokhrel}, R., {Gutermuth}, R.~A., {Betti}, S.~K., {et~al.} 2020, \apj, 896, 60

\bibitem[{{Portegies Zwart} {et~al.}(2010){Portegies Zwart}, {McMillan}, \&
  {Gieles}}]{PZ+2010review}
{Portegies Zwart}, S.~F., {McMillan}, S. L.~W., \& {Gieles}, M. 2010, \araa,
  48, 431

\bibitem[{{Rahner} {et~al.}(2019){Rahner}, {Pellegrini}, {Glover}, \&
  {Klessen}}]{Rahner+19}
{Rahner}, D., {Pellegrini}, E.~W., {Glover}, S. C.~O., \& {Klessen}, R.~S.
  2019, \mnras, 483, 2547

\bibitem[{{Reggiani} {et~al.}(2011){Reggiani}, {Robberto}, {Da Rio}, {Meyer},
  {Soderblom}, \& {Ricci}}]{Reggiani+2011}
{Reggiani}, M., {Robberto}, M., {Da Rio}, N., {et~al.} 2011, \aap, 534, A83

\bibitem[{{Schneider} {et~al.}(2015){Schneider}, {Bontemps}, {Girichidis},
  {Rayner}, {Motte}, {Andr{\'e}}, {Russeil}, {Abergel}, {Anderson},
  {Arzoumanian}, {Benedettini}, {Csengeri}, {Didelon}, {di}, {Griffin}, {Hill},
  {Klessen}, {Ossenkopf}, {Pezzuto}, {Rivera-Ingraham}, {Spinoglio},
  {Tremblin}, \& {Zavagno}}]{Schneider+15}
{Schneider}, N., {Bontemps}, S., {Girichidis}, P., {et~al.} 2015, \mnras, 453,
  L41

\bibitem[{{Shukirgaliyev}(2018)}]{BekPhDT18}
{Shukirgaliyev}, B. 2018, PhD thesis, Zentrum f{\"u}r Astronomie der
  Universit{\"a}t Heidelberg, Astronomisches Rechen-Institut, M{\"o}nchhofstr.
  12-14, D-69120 Heidelberg, Germany

\bibitem[{{Shukirgaliyev} {et~al.}(2017){Shukirgaliyev}, {Parmentier},
  {Berczik}, \& {Just}}]{Bek+17}
{Shukirgaliyev}, B., {Parmentier}, G., {Berczik}, P., \& {Just}, A. 2017, \aap,
  605, A119

\bibitem[{{Shukirgaliyev} {et~al.}(2019){Shukirgaliyev}, {Parmentier},
  {Berczik}, \& {Just}}]{Bek+19}
{Shukirgaliyev}, B., {Parmentier}, G., {Berczik}, P., \& {Just}, A. 2019,
  \mnras, 486, 1045

\bibitem[{{Shukirgaliyev} {et~al.}(2020){Shukirgaliyev}, {Parmentier},
  {Berczik}, \& {Just}}]{Bek+20}
{Shukirgaliyev}, B., {Parmentier}, G., {Berczik}, P., \& {Just}, A. 2020, in
  Star Clusters: From the Milky Way to the Early Universe, ed. A.~{Bragaglia},
  M.~{Davies}, A.~{Sills}, \& E.~{Vesperini}, Vol. 351, 507--511

\bibitem[{{Shukirgaliyev} {et~al.}(2018){Shukirgaliyev}, {Parmentier}, {Just},
  \& {Berczik}}]{Bek+18}
{Shukirgaliyev}, B., {Parmentier}, G., {Just}, A., \& {Berczik}, P. 2018, \apj,
  863, 171

\bibitem[{{Smith} {et~al.}(2011){Smith}, {Fellhauer}, {Goodwin}, \&
  {Assmann}}]{Smith+11}
{Smith}, R., {Fellhauer}, M., {Goodwin}, S., \& {Assmann}, P. 2011, \mnras,
  414, 3036

\bibitem[{{Teuben}(1995)}]{Teuben1995}
{Teuben}, P. 1995, in Astronomical Society of the Pacific Conference Series,
  Vol.~77, Astronomical Data Analysis Software and Systems IV, ed. R.~A.
  {Shaw}, H.~E. {Payne}, \& J.~J.~E. {Hayes}, 398

\bibitem[{{Tutukov}(1978)}]{Tutukov1978}
{Tutukov}, A.~V. 1978, \aap, 70, 57

\bibitem[{{Vasiliev}(2019)}]{agama}
{Vasiliev}, E. 2019, \mnras, 482, 1525

\bibitem[{{Verschueren}(1990)}]{Verschueren1990}
{Verschueren}, W. 1990, \aap, 234, 156

\bibitem[{{Verschueren} \& {David}(1989)}]{Verschueren+1989}
{Verschueren}, W. \& {David}, M. 1989, \aap, 219, 105

\bibitem[{{Wall} {et~al.}(2020){Wall}, {Mac Low}, {McMillan}, {Klessen},
  {Portegies Zwart}, \& {Pellegrino}}]{Wall+2020}
{Wall}, J.~E., {Mac Low}, M.-M., {McMillan}, S. L.~W., {et~al.} 2020, \apj,
  904, 192

\bibitem[{{Wall} {et~al.}(2019){Wall}, {McMillan}, {Mac Low}, {Klessen}, \&
  {Portegies Zwart}}]{Wall+2019}
{Wall}, J.~E., {McMillan}, S. L.~W., {Mac Low}, M.-M., {Klessen}, R.~S., \&
  {Portegies Zwart}, S. 2019, \apj, 887, 62

\end{thebibliography}

\end{document}